\documentstyle[twocolumn,epsf,eqsecnum,prb,aps]{revtex}

\begin{document}


\draft
\twocolumn[\hsize\textwidth\columnwidth\hsize\csname@twocolumnfalse%
\endcsname
\preprint{}
\title{Field dependence of the vortex structure in 
$d$-wave and $s$-wave superconductors}
\author{Masanori Ichioka, Akiko Hasegawa, and Kazushige Machida}
\address{Department of Physics, Okayama University, Okayama 700-8530, Japan}
\date{\today}
\maketitle
\begin{abstract}
We study the vortex structure and its field dependence within the framework 
of the quasi-classical  Eilenberger theory to find 
the difference between the $d_{x^2-y^2}$- and $s$-wave  pairings. 
We clarify the effect of the $d_{x^2-y^2}$-wave nature and 
the vortex lattice effect 
on the vortex structure of the pair potential, the internal field and 
the local density of states.  
The $d_{x^2-y^2}$-wave pairing introduces a fourfold-symmetric 
structure around each vortex core. 
With increasing field, their contribution becomes significant to the whole 
structure of the vortex lattice state, depending on the vortex lattice's  
configuration. 
It is reflected in the form factor of the internal field, 
which may be detected by small angle neutron scattering, or the 
 resonance line shape of $\mu$SR and NMR experiments. 
We also study the induced $s$- and $d_{xy}$-wave components around 
the vortex in $d_{x^2-y^2}$-wave superconductors. 
\end{abstract}

]

\narrowtext

\section{Introduction} 
\label{sec:introduction}

Much attention has been paid to the vortex structure 
in high-$T_c$ superconductors.  
Many researchers try to detect the $d_{x^2-y^2}$-wave nature 
of the superconductivity in the vortex structure of the mixed state.
It is necessary to clarify how the difference between 
the $d_{x^2-y^2}$-wave and conventional $s$-wave pairings appears 
in the mixed state. 
The point is how the vortex structure is affected by the anisotropy of the 
energy gap in the $d_{x^2-y^2}$-wave pairing, 
particularly by its nodal structure. 
It is known that the difference appears in the field ($H$) dependence of
the zero-energy density of states (DOS) $N(0)$. 
Volovik~\cite{Volovik} theoretically suggested that the dependence is
given by $N(0) \propto \sqrt{H}$  in the $d_{x^2-y^2}$-wave pairing 
and $N(0) \propto H$ in the $s$-wave pairing. 
The difference between the two pairings was indeed 
confirmed  by the self-consistent calculation of
the quasi-classical Eilenberger theory for the vortex lattice case, but  
the field dependence of $N(0)$ deviates from the exact $\sqrt{H}$- or 
$H$-linear behavior.~\cite{IchiokaL} 
In this paper, by using the quasi-classical method, 
we discuss other $d_{x^2-y^2}$-wave natures appearing  in the
vortex structure, such as the pair potential, the internal magnetic
field and the local density of states (LDOS). 

Low energy excitations, which dominantly govern the vortex structure, 
can be divided conceptually into three categories; 
(1) those from the continuum states associated with the nodal structure 
in the $d_{x^2-y^2}$-wave pairing, 
(2) the core excitations from the bound states localized in a vortex core, 
and (3) the quasiparticle transfer between vortices, 
i.e., the vortex lattice effect. 
While many calculations for the $d_{x^2-y^2}$-wave pairing 
such as Ref. \onlinecite{Volovik} take into account only item (1), 
which is valid near the lower critical field $H_{c1}$, 
items (2) and (3) are also indispensable when considering 
the vortex state in general. 
In order to help establish the general features of 
the mixed state in both the $d_{x^2-y^2}$- and $s$-wave cases, 
one needs to calculate the vortex structure while taking into account 
all three factors on an equal footing. 
Through these efforts we may gain a more valid and vivid picture of the 
vortex for the whole region of $H_{c1}<H<H_{c2}$ 
($H_{c2}$ is the upper critical field).

Detailed properties of the vortex structure, 
such as its field dependence,
attract much attention even in the conventional $s$-wave
superconductors. 
Several important means to probe the vortex structure are 
now available experimentally in various superconductors, including high-$T_c$ 
superconductors. 
Specific heat experiments~\cite{Moler,Fisher,Nohara,Hedo} 
inform us of low energy excitations. 
Muon spin resonance ($\mu$SR)\cite{SonierNbSe2,SonierYBCO} 
and small angle neutron scattering (SANS) investigate 
the internal field distribution of the vortex structure. 
Scanning tunneling microscopy (STM)\cite{Hess,Golubov,Maggio,Renner} 
directly observes the LDOS.  
These data are often analyzed within conventional phenomenological theories 
such as the Ginzburg-Landau (GL) theory or the London theory. 
The GL theory is, strictly speaking, valid only near the transition 
temperature $T_c$.\cite{Werthamer} 
As for the London theory, which is applied near $H_{c1}$, 
the simple cutoff procedure for the core radius is too rough an approximation 
to estimate the exact contribution of the vortex core.\cite{Brandt,Yaouanc} 
In this sense it is necessary to develop a microscopic theory in order 
to correctly analyze valuable experimental data. 

In this paper, we investigate various aspects of the vortex lattice structure 
based on the quasi-classical Eilenberger theory to clarify the difference 
in the vortex structure between the $d_{x^2-y^2}$- and $s$-wave pairings. 
The case of a magnetic field applied along the $c$ axis (or $z$ axis) 
is considered.   
We follow the method of Refs. \onlinecite{Klein} and 
\onlinecite{IchiokaS}, where the $s$-wave pairing case was investigated. 
This method can be applied in most regions of the mixed state, i.e., 
$0 < T < T_c$ and $H_{c1}<H<H_{c2}$. 
We calculate both the pair potential and the vector potential 
self-consistently in the vortex lattice case. 
Then we discuss the spatial variation of the current and
the internal field in detail. 
As for the single vortex case, the fourfold-symmetric vortex core 
structure of the $d_{x^2-y^2}$-wave pairing was shown 
in Ref. \onlinecite{IchiokaD} by the quasi-classical calculation. 
In the present paper, where we examine the vortex lattice case, we can see how 
this $d_{x^2-y^2}$-wave nature and the vortex lattice effect 
affect each other. 
And we also study the field dependence of the vortex structure.  

The rest of this paper is organized as follows. 
In Sec. \ref{sec:formulation}, we describe the method of the 
quasi-classical calculation. 
In Sec. \ref{sec:configulation}, we estimate the free energy for several
configurations of the vortex lattice. 
There, the square lattice has lower free energy than the
conventional triangular lattice at higher fields. 
In Sec.  \ref{sec:square}, 
we study the field dependence of the vortex structure 
(the pair potential, the current and the internal field) in the square
lattice case. 
The form factor and the distribution function of the internal field are 
discussed in connection with the SANS and $\mu$SR experiments. 
The vortex structure in the triangular lattice case is briefly discussed 
in Sec. \ref{sec:triangular}. 
We study the LDOS structure and the field dependence of the 
spatially averaged DOS in Sec. \ref{sec:dos}, 
and the structure of the induced order parameter of 
the other symmetry in Sec. \ref{sec:induced}. 
A summary and discussions are given in Sec. \ref{sec:summary}.

\section{quasi-classical Eilenberger theory}
\label{sec:formulation}

Our calculations are performed following the method of
Refs. \onlinecite{Klein} and \onlinecite{IchiokaS}. 
We consider the case of the clean limit and the cylindrical Fermi surface. 
These are appropriate to high-$T_c$ superconductors 
and low-dimensional organic superconductors. 

First, we obtain the pair potential and the vector potential self-consistently 
by solving the Eilenberger equation in the Matsubara 
frequency $\omega_n=(2n+1)\pi T$. 
We consider the quasi-classical Green functions 
$g(i\omega_n,\theta,{\bf r})$, $f(i\omega_n,\theta,{\bf r})$ and 
$f^\dagger(i\omega_n,\theta,{\bf r})$, 
where ${\bf r}$ is the center of mass coordinate of the Cooper pair. 
The direction of the relative momentum of the Cooper pair, 
$\hat{\bf k}={\bf k}/|{\bf k}|$, is denoted by an angle $\theta$ 
measured from the $x$ axis. 
The Eilenberger equation is given by~\cite{Eilenberger}
\begin{eqnarray} &&
\Bigl\{ \omega_n +{i \over 2} {\bf v}_{\rm F}\cdot 
\Bigl(\frac{\nabla}{i}+ \frac{2\pi}{\phi_0} {\bf A}({\bf r}) \Bigr) \Bigr\} 
f(i\omega_n,\theta,{\bf r})  
\nonumber \\ &&
= \Delta(\theta,{\bf r}) g(i\omega_n,\theta,{\bf r}) ,
\label{eq:eil-1} \\  && 
\Bigl\{ \omega_n -{i \over 2} {\bf v}_{\rm F}\cdot
\Bigl(\frac{\nabla}{i}-\frac{2\pi}{\phi_0} {\bf A}({\bf r}) \Bigr) \Bigr\}
f^\dagger(i\omega_n,\theta,{\bf r}) 
\nonumber \\ &&
= \Delta^\ast(\theta,{\bf r}) g(i\omega_n,\theta,{\bf r}),
\label{eq:eil-2} \\  &&
g(i\omega_n,\theta,{\bf r}) =[1- f(i\omega_n,\theta,{\bf r})
    f^\dagger(i\omega_n,\theta,{\bf r}) ]^{1/2}, 
\label{eq:eil-3} 
\end{eqnarray}
where ${\rm Re} g(i\omega_n,\theta,{\bf r}) > 0$ and 
${\bf v}_{\rm F}=v_{\rm F}\hat{\bf k}$ is the Fermi velocity.
The vector potential 
is written as 
${\bf A}({\bf r})=\frac{1}{2} {\bf H} \times {\bf r} + {\bf a}({\bf r})$
in the symmetric gauge, where ${\bf H}=(0,0,H)$ is a  
uniform field and ${\bf a}({\bf r})$ is related to the internal field 
${\bf h}({\bf r})=(0,0,h({\bf r}))$,  
where ${\bf h}({\bf r})=\nabla\times {\bf a}({\bf r})$.
As for the pair potential 
$\Delta(\theta,{\bf r})=\Delta({\bf r})\phi(\theta)$, 
we set $\phi(\theta)=\sqrt{2}\cos2(\theta-\theta_0)$ for 
the $d_{x^2-y^2}$-wave pairing and  $\phi(\theta)=1$ for the $s$-wave pairing. 
Here, $\theta_0$ is the angle between the $x$ axis and the $a$ axis of
the crystal coordinate. 
The self-consistent conditions for $\Delta({\bf r})$ and ${\bf a}({\bf r})$ 
are given as 
\begin{equation}
\Delta(\theta,{\bf r})= N_0 
2 \pi T \sum_{\omega_n>0} \int_0^{2\pi}{d\theta' \over 2\pi}
V(\theta',\theta)f(i\omega_n,\theta',{\bf r}) ,
\label{eq:self-d}
\end{equation}
\begin{equation}
\nabla\times\nabla\times{\bf a}({\bf r})
= - \frac{\pi\phi_0}{\kappa^2 \Delta_0 \xi_0^3} 2\pi T \sum_{\omega_n>0} 
\int_0^{2\pi}{d\theta \over 2\pi}{\hat{\bf k} \over i}
g(i \omega_n,\theta,{\bf r}) , 
\label{eq:self-a}
\end{equation}
where $N_0$ is the density of states at the Fermi surface, 
$V(\theta',\theta)=\bar{V}\phi(\theta')\phi(\theta)$ is 
the pairing interaction, 
$\kappa=(7 \zeta(3)/72)^{1/2}(\Delta_0/ T_c)\kappa_{\rm BCS}$  
with Rieman's zeta function $\zeta(3)$. 
$\kappa_{\rm BCS}$ is the GL parameter in the BCS theory, and 
$\Delta_0$ is the uniform gap at $T=0$.  
We set the energy cutoff $\omega_c=20 T_c$.
In the following calculations, 
energies and lengths are measured in units of $\Delta_0$ 
and $\xi_0=v_{\rm F}/\Delta_0 =\pi \xi_{\rm BCS}$ 
($\xi_{\rm BCS}$ is the BCS coherence length), respectively.

By solving Eqs. (\ref{eq:eil-1})-(\ref{eq:eil-3}) by the so-called explosion 
method~\cite{Klein,IchiokaS} 
under $\Delta({\bf r})$ and ${\bf a}({\bf r})$ of the vortex lattice case, 
we estimate the quasi-classical Green functions at $40\times 40$ discretized 
points in a unit cell of the vortex lattice. 
We obtain a new $\Delta({\bf r})$ and ${\bf a}({\bf r})$ from 
Eqs. (\ref{eq:self-d}) and (\ref{eq:self-a}), and use them at the next step 
in calculating of Eqs. (\ref{eq:eil-1})-(\ref{eq:eil-3}). 
This iteration procedure is repeated until a sufficiently 
self-consistent solution is obtained. 
We use the material parameters appropriate to YBCO, i.e., 
$\xi_{\rm BCS}$=16\AA \   and $\kappa_{\rm BCS}$=100. 
There, $H_{c2}$=66.7 Tesla in the $s$-wave pairing and 93.2 Tesla in 
the $d_{x^2-y^2}$-wave pairing for $T/T_c=0.5$.~\cite{Sugiyama} 
To study the field dependence, the calculations are done for various 
fields at a fixed temperature $T/T_c=0.5$. 
The spatial variation of the internal field and the current 
${\bf J}({\bf r})=(c/4\pi)\nabla\times{\bf h}({\bf r})$ is calculated 
from ${\bf a}({\bf r})$.

Next, we calculate the LDOS for energy $E$ by  
\begin{equation}
N(E,{\bf r})= N_0 \int_0^{2\pi} \frac{d\theta}{2\pi} 
{\rm Re}\ g(i\omega_n \rightarrow E+i\eta,\theta,{\bf r}) .  
\label{eq:ldos} 
\end{equation}
To obtain $ g(i\omega_n \rightarrow E+i\eta,\theta,{\bf r})$, 
we solve Eqs. (\ref{eq:eil-1})-(\ref{eq:eil-3}) for $\eta-iE$ instead 
of $\omega_n$ using the self-consistently obtained $\Delta({\bf r})$
and ${\bf a}({\bf r})$. 
The DOS is given by 
\begin{equation}
N(E)=\langle N(E,{\bf r}) \rangle.  
\label{eq:dos}
\end{equation}
Here, 
$\langle\cdots\rangle =  \int_{\rm unit \ cell} d{\bf r} (\cdots)/S$ 
means the spatial average, where $S$ is the area of a unit cell. 

Free energy $F$ is calculated by~\cite{Klein,Eilenberger} 
\begin{eqnarray}
\frac{F}{N_0 \Delta_0^2 }
&=&
 \kappa^2 \frac{\langle h({\bf r})^2 \rangle}{(\phi_0/\xi_0^2)^2} 
+ \frac{\langle |\Delta({\bf r})|^2 \rangle}{N_0 \bar{V} \Delta_0^2}
\nonumber \\ && 
-\frac{2 \pi T}{\Delta_0^2}\sum_{\omega_n > 0}\int_0^{2\pi} 
\frac{d \theta}{2 \pi}\langle I(\omega_n,\theta,{\bf r}) \rangle  
\label{eq:free-energy}
\end{eqnarray}
with 
\begin{eqnarray} &&
I(\omega_n,\theta,{\bf r})
=
\Delta^\ast(\theta,{\bf r}) f
+\Delta(\theta,{\bf r}) f^\dagger
\nonumber \\ &&
-( 1-g ) 
\Bigl[ \frac{1}{f}
\Bigl\{ \omega_n +{i \over 2} {\bf v}_{\rm F}\cdot
\Bigl(\frac{\nabla}{i}+ \frac{2\pi}{\phi_0} {\bf A}({\bf r}) \Bigr) \Bigr\}f
\nonumber \\ &&
+\frac{1}{f^\dagger}
\Bigl\{ \omega_n -{i \over 2} {\bf v}_{\rm F}\cdot
\Bigl(\frac{\nabla}{i}-\frac{2\pi}{\phi_0} {\bf A}({\bf r}) \Bigr) \Bigr\}
f^\dagger\Bigr] 
\label{eq:free-energy2} \\ &&
=\frac{\Delta^\ast(\theta,{\bf r}) f +\Delta(\theta,{\bf r}) f^\dagger}{1+g}, 
\label{eq:free-energy3}
\end{eqnarray}
where $g$, $f$ and $f^\dagger$ mean $g(i\omega_n,\theta,{\bf r}) $, 
$f(i\omega_n,\theta,{\bf r}) $ and  $f^\dagger(i\omega_n,\theta,{\bf r}) $, 
respectively. 
We use Eqs. (\ref{eq:eil-1})-(\ref{eq:eil-3}) to obtain 
Eq. (\ref{eq:free-energy3}). 

\section{vortex lattice configuration}
\label{sec:configulation}

We have to fix the shape and the orientation of the vortex lattice
before calculation. 
The shape of the vortex lattice is characterized
by the unit vectors ${\bf r}_1=(a_x,0,0)$ and ${\bf r}_2=(\zeta a_x,a_y,0)$ 
where $a_xa_y=\phi_0/H$ and where the $x$ axis is set in 
the ${\bf r}_1$ direction. 
Since we consider the case $|{\bf r}_1|=|{\bf r}_2|$, 
we can write ${\bf r}_2=(a_x \cos\theta_L, a_x \sin\theta_L)$, 
where $\theta_L$ is the angle between ${\bf r}_1$ and ${\bf r}_2$. 

In this paper, we study the contributions of the $d_{x^2-y^2}$-wave
effect and the vortex lattice effect on the vortex structure. 
As the $d_{x^2-y^2}$-wave nature reflects the node direction of the
energy gap, its directional dependence is determined by the angle from the 
crystal coordinate. 
To investigate the $d_{x^2-y^2}$-wave nature, we compare the
vortex structure along the $a$ axis and $b$ axis directions (we denote
these directions as ``the $0^\circ$ directions'' in the following) and that
along the $45^\circ$ directions from the $a$ axis (we denote these as
``the $45^\circ$ directions''). 
On the other hand, to study the vortex lattice effect, we compare the
vortex structure along the nearest neighbor (NN) vortex direction  
and that along the next nearest neighbor (NNN) vortex direction. 
We denote the former direction as the ``NN direction'', and the latter
direction as the ``NNN direction''. 
The orientation of the vortex lattice, i.e., the relative angle between 
the ``$0^\circ$ directions''  and the ``NN direction'' is 
important in considering the vortex lattice structure in the
$d_{x^2-y^2}$-wave pairing. 
The orientation is characterized by $\theta_0$. 

\begin{figure}
\leavevmode
\epsfxsize=6.0cm
\epsfbox{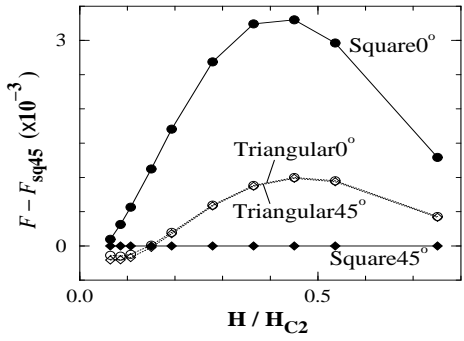}
\vspace{0.2cm}
\caption{
Field dependence of the free energy difference $F-F_{\rm sq45}$, which
 is normalized by $N_0 \Delta_0^2$. 
The square vortex lattice case and the triangular lattice case are presented 
for $\theta_0=45^\circ$ and $0^\circ$. 
}
\label{fig:free-energy}
\end{figure}

In an isotropic $s$-wave superconductor, the shape of the vortex lattice 
is a $60^\circ$ triangular lattice. 
However, in a $d_{x^2-y^2}$-wave superconductor, the shape may 
deviate from a triangular lattice. 
The deviation comes from the fourfold-symmetric vortex structure 
around the vortex core, which reflects the $d_{x^2-y^2}$-wave symmetry 
of the pair potential in the ${\bf k}$ space 
(i.e., $\hat k_x^2 -\hat k_y^2$). 
To determine the vortex lattice configuration of the equilibrium state, 
we compare $F$ of Eq. (\ref{eq:free-energy}) for the various vortex
lattice configurations, and find the state having minimum $F$. 
As our quasi-classical calculation needs a lot of time to compute, we
cannot check all possible configurations of shape and orientation. 
So, we only compare the square lattice case ($\theta_L=90^\circ$) 
and the triangular lattice case ($\theta_L=60^\circ$) 
in the $d_{x^2-y^2}$-wave pairing. 
There, we consider the two orientation cases $\theta_0=45^\circ$ and
$0^\circ$. 
For $\theta_0=45^\circ$ ($\theta_0=0^\circ$), one of the NN vortices is 
located in the $45^\circ$ ($0^\circ$) direction. 
The field dependence of $F$ is shown in Fig. \ref{fig:free-energy}.  
There, we present the difference $F-F_{\rm sq45}$, where $F_{\rm sq45}$
is $F$ for the case of the square lattice with $\theta_0=45^\circ$. 
The square lattice with $\theta_0=45^\circ$ has the smallest $F$ at a higher
field $H/H_{c2}>0.15$. 
At a lower field, the triangular lattice case has the smallest $F$. 
So, the vortex lattice configuration changes from the triangular 
lattice to the square lattice with $\theta_0=45^\circ$ as the field increases. 
We note that there may be a stable oblique lattice 
between the triangular lattice and the square lattice. 
An estimate including the oblique lattice 
by the quasi-classical theory belongs to a future work. 
As the square lattice with $\theta_0=0^\circ$ has a higher $F$, it is an
unstable state. 
As for the orientation of the triangular lattice, the case 
$\theta_0=45^\circ$ has a smaller $F$ than the case $\theta_0=0^\circ$, 
but the difference is very small. 

Some theoretical calculations considered the vortex lattice configuration 
near $H_{c2}$,~\cite{IchiokaGL,Won,DeWilde}  
or based on the London theory.\cite{Kogan,FranzL,Amin}
These theories suggested that the vortex lattice configuration  
gradually changes from the triangular lattice to the square lattice 
with increasing field, and the oblique lattice is 
realized at a lower field before settling into the square lattice. 
The square lattice configuration with
$\theta_0=45^\circ$ has lower free energy than that of the
conventional $60^\circ$ triangular 
lattice over a wide region of higher fields and lower temperatures  
in  the mixed state for the $d_{x^2-y^2}$-wave pairing. 
This is consistent with the estimate of Fig. \ref{fig:free-energy}. 
An STM experiment on ${\rm YBa_2Cu_3O_{7-\delta}}$ (YBCO) also 
suggested the square vortex lattice configuration 
in the $d_{x^2-y^2}$-wave pairing.~\cite{Maggio}  

\section{vortex structure in the square lattice case}
\label{sec:square} 

As noted in the previous section, 
the stable configuration of the vortex lattice at higher field 
is the square lattice with $\theta_0=45^\circ$.  
We fix the shape of the vortex lattice as the square lattice in order 
to study the field dependence of the vortex structure.  
To clarify the roles of the $d_{x^2-y^2}$-wave effect and the vortex
lattice effect, we present the results of the unstable orientation case 
$\theta_0=0^\circ$ in addition to the stable orientation case
$\theta_0=45^\circ$ in the  $d_{x^2-y^2}$-wave pairing. 
We denote the former as ``the $d0^\circ$ case'', and the latter as 
``the $d45^\circ$ case''. 
The $s$-wave pairing case is also calculated for the square lattice 
to clarify the effect of low lying excitations associated with the gap 
anisotropy. 
The comparison among these cases ($s$-wave case, $d45^\circ$ case 
and $d0^\circ$ case) helps us to understand the contributions of 
the $d_{x^2-y^2}$-wave nature and the vortex lattice effect.

\subsection{Amplitude of the pair potential}

\begin{figure*}
\leavevmode
\epsfxsize=12.0cm
\epsfbox{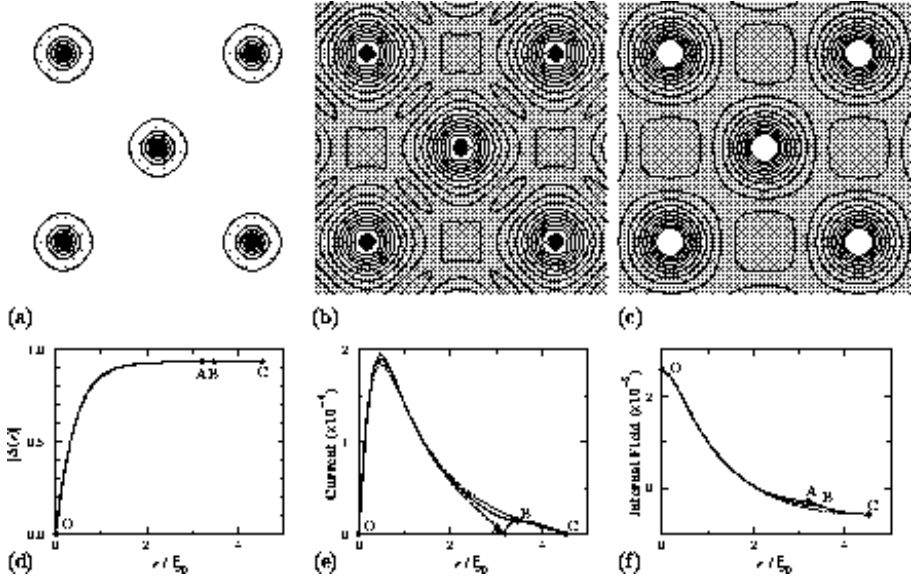}
\vspace{0.2cm}
\caption{
Spatial variation of the vortex structure at low field $H
/H_{c2}$=0.021 for the $d45^\circ$ case. 
The $a$ axis (i.e., $0^\circ$ direction) is horizontal. 
(a) Amplitude of the pair potential. 
The contour lines of 
$|\Delta({\bf r})|/\Delta_0$ are plotted. 
From the center, 
the lines are for $|\Delta({\bf r})|/\Delta_0$=0.1, 0.2,..., 0.9.
Contour lines show a fourfold-symmetric shape around each vortex. 
(b) Amplitude of the current. Contour lines of
$|{\bf J}({\bf r})|/(c \phi_0 /4 \pi \xi_0)$ are plotted. 
There are four small peaks around each vortex. 
(c) Internal field distribution. 
Contour lines of $h({\bf r})/(\phi_0/\xi_0^2)$  are plotted. 
(d)-(f) are, respectively,  the profiles of (a)-(c) 
along the lines OA (solid line), OB (dotted line), OC (dot-dashed line) 
and AC (thick line) of Fig. \protect{\ref{fig:line}}. 
}
\label{fig:lowH}
\end{figure*}

In Fig. \ref{fig:lowH}, we show the spatial variation of 
the vortex structure at low field $H/H_{c2}=0.021$ for  
the $d45^\circ$ case.
We show their contour lines and their profile. 
The profile is presented along the lines OA, OB, OC and AC 
shown in Fig. \ref{fig:line}.
The line OA (OC) is the radius in the NN (NNN) direction, and the 
line AC is along the boundary of the Wigner-Seitz cell of the vortex lattice. 

The amplitude $|\Delta({\bf r})|$ is presented in Figs. \ref{fig:lowH} 
(a) and (d).
At low field, the vortex core region occupies a small part of a unit
cell of the vortex lattice. 
There, $|\Delta({\bf r})|$ approaches a constant value ($0.95 \Delta_0$
for $T=0.5T_c$) at a few $\xi_0$ away from the vortex center, 
as presented in Fig. \ref{fig:lowH} (d). 
The $d_{x^2-y^2}$-wave nature appears in the shape of the contour lines of 
$|\Delta({\bf r})|$.  
In the $s$-wave pairing, the contour lines are circular around each
vortex center at low field. 
In the $d_{x^2-y^2}$-wave pairing, as shown in Fig. \ref{fig:lowH} (a), 
the contour lines become a fourfold-symmetric shape, 
especially for $0.5 \Delta_0 \le |\Delta({\bf r})| \le 0.9 \Delta_0$. 
There $|\Delta({\bf r})|$ is suppressed around the vortex core 
in the $0^\circ$ directions compared with the $45^\circ$ directions. 
This structure is consistent with that obtained by the single vortex 
calculation.~\cite{IchiokaD}

Figure \ref{fig:damp-high} shows $|\Delta({\bf r})|$ 
at a higher field $H/H_{c2}=0.54$ for three cases.  
At high field, the vortex core region occupies a large part of a unit 
cell. 
So, $|\Delta({\bf r})|$ does not reduce to a uniform value even in
the boundary region, as shown in Figs. \ref{fig:damp-high} (d)-(f). 
The influence of the vortex lattice effect on $|\Delta({\bf r})|$ 
can be seen in Figs. \ref{fig:damp-high} (a) and (d) for 
the $s$-wave pairing case. 
While the vortex core has a circular symmetric structure in the inner
region of the core ($|\Delta({\bf r})| \le 0.5 \Delta_0$), 
$|\Delta({\bf r})|$ shows directional dependence 
in the outer region,  reflecting the NN vortices. 
There, $|\Delta({\bf r})|$ is suppressed in the NN directions (line OA)
compared with the NNN directions (line OC). 
In the $d_{x^2-y^2}$-wave pairing, the contribution of
the fourfold-symmetric core structure appears in addition to the vortex 
lattice effect, depending on the vortex lattice's configuration,
and especially on its orientation. 
We demonstrate this by showing two orientation cases $d45^\circ$ 
[Figs. \ref{fig:damp-high} (b) and (e)] and $d0^\circ$ 
[Figs. \ref{fig:damp-high} (c) and (f)]. 
In both cases, the inner region of the vortex core shows a 
fourfold-symmetric structure, seen also in the low field case 
shown in Fig. \ref{fig:lowH} (a). 
However, the difference between the two orientation cases appears in the
outer region of the vortex core. 
Due to the $d_{x^2-y^2}$-wave nature, $|\Delta({\bf r})|$ is suppressed
in the $0^\circ$ directions. 
Because of the vortex lattice effect, $|\Delta({\bf r})|$ is suppressed
in the NN directions. 
When these two directions disagree as in Fig. \ref{fig:damp-high} (b), 
these two suppressions cancel each other. 
The difference between $|\Delta({\bf r})|$ in the NN direction
(line OA) and $|\Delta({\bf r})|$ in the NNN direction (line OC) is smeared. 
Along the boundary line AC, $|\Delta({\bf r})|$ is almost constant. 
On the other hand, when the $0^\circ$ direction and the NN direction agree 
as in Fig. \ref{fig:damp-high} (c), the suppressions due to the vortex 
lattice effect and the $d_{x^2-y^2}$-wave effect enhance each other. 
$|\Delta({\bf r})|$ is largely suppressed in the $0^\circ$ 
directions (i.e., the line OA in the NN direction) compared to 
$|\Delta({\bf r})|$ in the $45^\circ$ direction 
(i.e., the line OC in the NNN direction). 
The result shown in  Fig. \ref{fig:free-energy}, which shows that the
$d45^\circ$ case is stable at higher field and the $d0^\circ$ case is
unstable,  may be related to the 
difference in the vortex structure shown in 
Figs. \ref{fig:damp-high} (b) and (c). 

Next, we study the field dependence of $|\Delta({\bf r})|$. 
We define $\Delta_{\rm NN}$ ($\Delta_{\rm NNN}$) as the maximum of
$|\Delta({\bf r})|$ along the line OA (OC) in the NN (NNN) direction. 
The field dependence of $\Delta_{\rm NN}$ and $\Delta_{\rm NNN}$ is
presented in Fig. \ref{fig:damp-max} for the $s$-wave (a) and the 
$d_{x^2-y^2}$-wave (b) pairings. 
They decrease to 0 as $H_{c2}$ is approached. 
The difference between $\Delta_{\rm NN}$ and $\Delta_{\rm NNN}$, which
reflects the vortex lattice effect, appears when $H/H_{c2} \ge 0.1$. 
The difference is small for the $d45^\circ$ case, but large for
the $d0^\circ$ case, as discussed above. 

The field dependence of the core radius is shown in
Fig. \ref{fig:radius}. 
The radius $\xi_1$ is defined from the initial slope of the pair
potential by setting as $|\Delta({\bf r})|=\Delta_{\rm NN} r/ \xi_1$ at
the vortex center. 
As $H$ increases, the radius $\xi_1$ decreases similarly for both 
the $s$-wave and the $d45^\circ$ cases.  
In the unstable $d0^\circ$ case, the decrease of $\xi_1$ is saturated
at a higher field compared to the other cases. 

\subsection{Internal current distribution} 

The spatial variation of the current $|{\bf J}({\bf r})|$ at low field
$H/H_{c2}=0.021$ is shown in Figs. \ref{fig:lowH} (b) and (e) 
for the $d45^\circ$ case. 
There, $|{\bf J}({\bf r})|$ has four peaks around each vortex core in
the $45^\circ$ directions at $r \sim 0.5 \xi_0$. 
The four peaks are $d_{x^2-y^2}$-wave nature, and are consistent with 
the single vortex calculation.~\cite{IchiokaD}  
In the $s$-wave pairing at low field, 
$|{\bf J}({\bf r})|$ distributes circularly
around each vortex without four peaks. 
Outside the vortex core region ($r \ge \xi_0$), there is no clear difference
between the $s$- and $d_{x^2-y^2}$-wave pairings in the current
distribution.  

Figure \ref{fig:jamp-high} shows $|{\bf J}({\bf r})|$ at high field
$H/H_{c2}=0.54$, which corresponds to the case of Fig. \ref{fig:damp-high}. 
In the $s$-wave pairing case of Fig. \ref{fig:jamp-high} (a), there are
four peaks around each vortex core in the NN vortex directions. 
These peaks are due to the vortex lattice effect and enhance the four peaks 
of the $d_{x^2-y^2}$-wave nature in the $45^\circ$ directions 
for the $d45^\circ$ case [Fig. \ref{fig:jamp-high} (b)], but
smear them for the $d0^\circ$ case [Fig. \ref{fig:jamp-high} (c)]. 

We can also define the core radius by the maximum current. 
The core radius $\xi_2$ is defined as the radius where 
$|{\bf J}({\bf r})|$ is at a maximum in the NN vortex direction. 
The field dependence of $\xi_2$ is presented in Fig. \ref{fig:radius} (b). 
The radius $\xi_2$ decreases similarly to $\xi_1$ as shown in 
Fig. \ref{fig:radius} (a). 
From both definitions of $\xi_1$ and $\xi_2$ (i.e., the initial slope of
$|\Delta({\bf r})|$ and the maximum current), we obtain a similar
dependence on $H$. 
The shrinkage of the core radius was also reported in 
STM and $\mu$SR experiments.~\cite{Golubov,SonierNbSe2} 
There, the results of the field dependence were analyzed in the dirty
limit theory by the Wigner-Seitz method (i.e., a circular unit cell is
used instead of the periodic boundary condition of the vortex lattice).
In our study, we obtain the field dependence in the clean limit, 
where we treat the periodic boundary condition exactly. 

The radius is also presented for the triangular vortex lattice of the
$s$-wave pairing[Fig. \ref{fig:radius}]. 
The field dependence is almost the same as in the square lattice case of
the $s$-wave pairing. 
We confirm that the core radius of the $d_{x^2-y^2}$-wave pairing 
shows a similar decrease to the triangular vortex lattice 
case[Fig. \ref{fig:radius}]. 
When we calculate the radius for a smaller $\kappa_{\rm BCS}$ 
($\kappa_{\rm BCS}=7$) 
in the $s$-wave pairing to study the case of ${\rm NbSe_2}$, 
a similar decrease is obtained. 
So, the decrease of the core radius occurs over a wide range of
$\kappa_{\rm BCS}$. 

\subsection{Internal field distribution}

Reflecting the four peaks of $|{\bf J}({\bf r})|$ in the $45^\circ$
directions, the internal field distribution has a fourfold-symmetric
structure around each vortex core, 
as presented by the single vortex calculation.\cite{IchiokaD}
$h({\bf r})$ extends toward the $0^\circ$ directions  
in which the screening current is weak. 
However, this fourfold-symmetric structure is restricted to the vortex core
region, since the $d_{x^2-y^2}$-wave nature is compensated by 
${\bf J}({\bf r})$ outside the core. 
Outside the core region, $h({\bf r})$ immediately reduces to 
a circular structure. 
This compensation is seen in the single vortex 
case,\cite{IchiokaD,Enomoto} 
but it also appears in the vortex lattice case at low field.
In Fig. \ref{fig:lowH} (c) and (f), 
we show the spatial variation of $h({\bf r})$
at low field $H/H_{c2}=0.021$ for the $d45^\circ$ case. 
In the $d_{x^2-y^2}$-wave pairing, the contour lines of $h({\bf r})$ in
the vortex core region are distorted from the circle of 
the $s$-wave pairing case into a fourfold symmetry. 
The distortion can be seen within the small core region as shown in  
Fig. \ref{fig:lowH} (c).  
Outside the core region, $h({\bf r})$ recovers the circular contour
lines seen in the $s$-wave case. 
So the $d_{x^2-y^2}$-wave nature scarcely affects the internal field
distribution at low field. 
In the $s$-wave pairing case, $h({\bf r})$ has the same structure
as that shown in Fig. \ref{fig:lowH} (c) except for the small region of the
vortex core. 

\begin{figure}
\leavevmode
\epsfxsize=3.5cm
\epsfbox{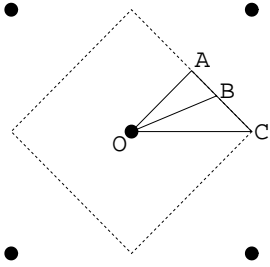}
\vspace{0.2cm}
\caption{
Square vortex lattice configuration in a contour plot 
for the $s$-wave and the $d45^\circ$ cases. 
The $a$ axis (i.e., $0^\circ$ direction) is horizontal. 
For the $d0^\circ$ case, this configuration is rotated by $45^\circ$. 
The vortex centers are shown by solid circles.
A square enclosed by dashed lines represents the Wigner-Seitz cell of the 
vortex lattice. 
In Figs. \protect{\ref{fig:lowH}}, \protect{\ref{fig:damp-high}}, 
\protect{\ref{fig:jamp-high}} and \protect{\ref{fig:h-high}}, 
the profile of the vortex structure is presented along the lines 
OA, OB, OC and AC.  
} 
\label{fig:line} 
\end{figure}

\begin{figure*}
\leavevmode
\epsfxsize=12.0cm
\epsfbox{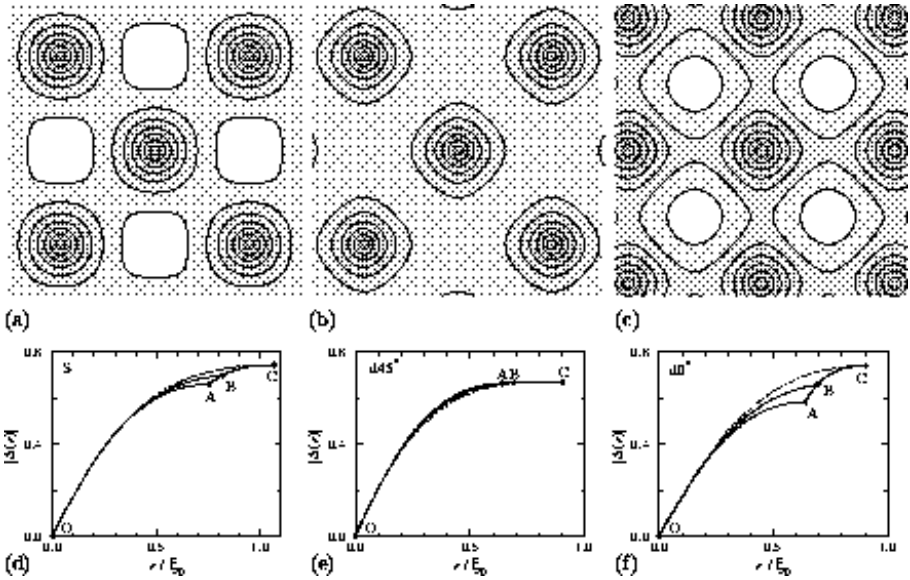}
\vspace{0.2cm}
\caption{
Spatial variation of the pair potential at high field
$H/H_{c2}$=0.54 for the $s$-wave (a), the $d45^\circ$ (b) and 
the $d0^\circ$ (c) cases.
The contour lines of $|\Delta({\bf r})|/\Delta_0$ are plotted.
From the center, 
the lines are for $|\Delta({\bf r})|/\Delta_0$=0.1, 0.2,..., 0.7.
The $a$ axis (i.e., $0^\circ$ direction) is horizontal. 
(d)-(f) are, respectively,  profiles of (a)-(c)
along the lines OA, OB, OC and AC of Fig. \protect{\ref{fig:line}}.
}
\label{fig:damp-high}
\end{figure*}

\begin{figure}
\leavevmode
\epsfxsize=8.0cm
\epsfbox{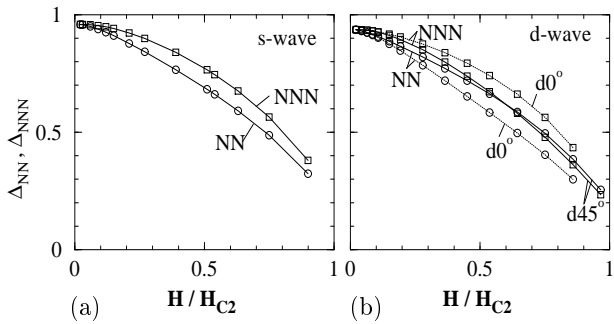}
\vspace{0.2cm}
\caption{
Field dependence of $\Delta_{\rm NN}$ and $\Delta_{\rm NNN}$, 
which are normalized by $\Delta_0$. 
We define $\Delta_{\rm NN}$ ($\Delta_{\rm NNN}$) as the maximum of 
$|\Delta({\bf r})|$ along the NN (NNN) direction. 
(a) $s$-wave case. (b) $d45^\circ$ and $d0^\circ$ cases.
}
\label{fig:damp-max}
\end{figure}

\begin{figure}
\leavevmode
\epsfxsize=8.0cm
\epsfbox{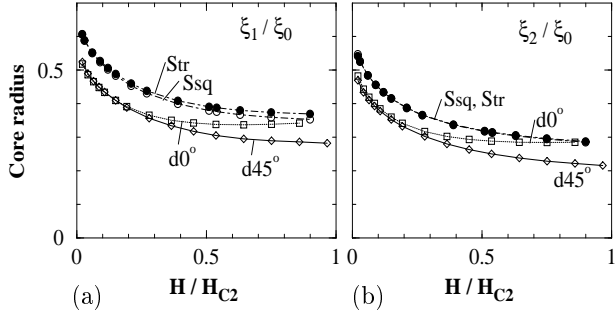}
\vspace{0.2cm}
\caption{
Field dependence of the vortex core radius $\xi_1$ (a) and $\xi_2$ (b) 
for the $s$-wave ($\circ$), the $d45^\circ$ ($\diamond$) and the 
$d0^\circ$ ($\Box$) cases in the square lattice. 
They are normalized by $\xi_0$. 
The radius $\xi_1$ is defined from the initial slope of the pair potential 
by setting as $|\Delta({\bf r})|=\Delta_{\rm NN}r/\xi_1$. 
The radius $\xi_2$ is defined from the maximum of the screening current 
$|{\bf J}({\bf r})|$.  
Both $\xi_1$ and $\xi_2$ show a similar field dependence. 
The triangular lattice case of the $s$-wave pairing is also 
plotted ($\bullet$). 
Its field dependence is almost the same as the square lattice case of
the $s$-wave pairing. 
}
\label{fig:radius}
\end{figure}

\begin{figure*}
\leavevmode
\epsfxsize=12.0cm
\epsfbox{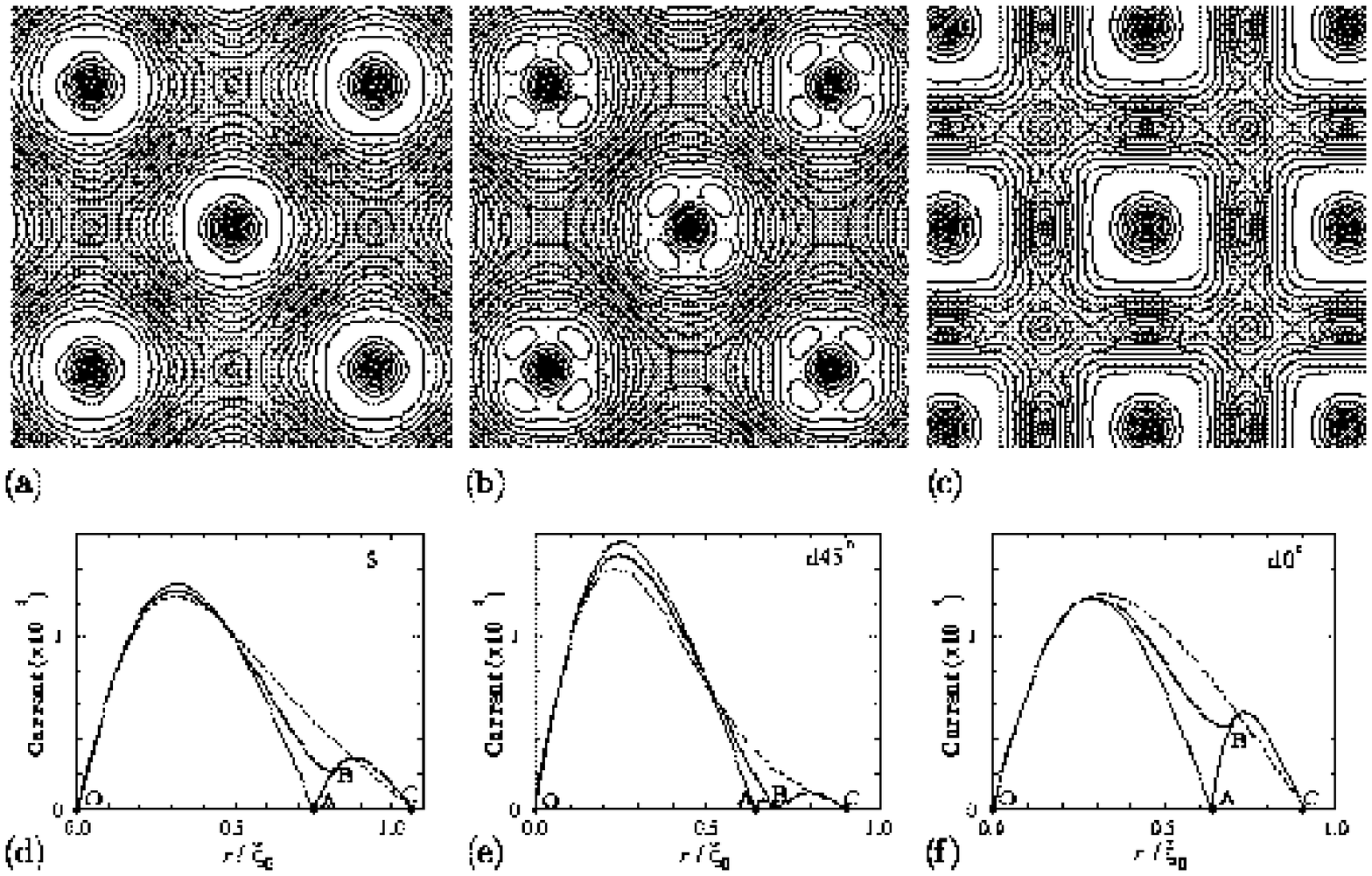}
\vspace{0.2cm}
\caption{
Spatial variation of the internal current at high field
$H/H_{c2}$=0.54 for the $s$-wave (a), the $d45^\circ$ (b) and
the $d0^\circ$ (c) cases.
Contour lines of $|{\bf J}({\bf r})|/(c \phi_0/ 4 \pi \xi_0)$ are plotted.
(d)-(f) are, respectively,  profiles of (a)-(c).
}
\label{fig:jamp-high}
\end{figure*}

\begin{figure*}
\leavevmode
\epsfxsize=12.0cm
\epsfbox{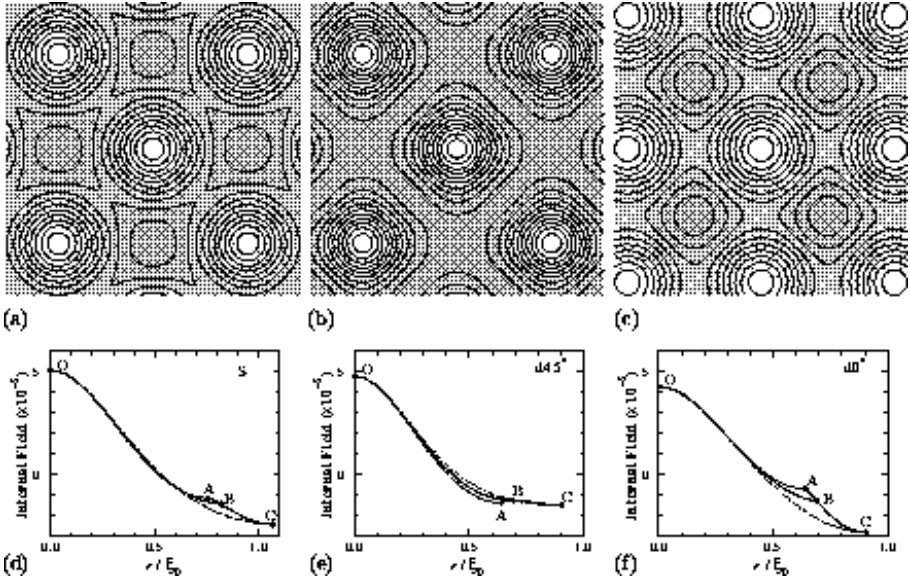}
\vspace{0.2cm}
\caption{
Spatial variation of the internal magnetic field at high field
$H/H_{c2}$=0.54 for the $s$-wave (a), the $d45^\circ$ (b) and
the $d0^\circ$ (c) cases.
Contour lines of $h({\bf r})/(\phi_0/\xi_0^2)$ are plotted.
(d)-(f) are, respectively,  profiles of (a)-(c). 
}
\label{fig:h-high}
\end{figure*}

Figure \ref{fig:h-high} shows $h({\bf r})$ at high field
$H/H_{c2}$=0.54, which corresponds to the case shown in 
Fig. \ref{fig:jamp-high}. 
In the $s$-wave pairing, $h({\bf r})$ is lowest at the boundary of the 
NNN direction (i.e., the point C in Fig. \ref{fig:line}). 
There is a saddle point in $h({\bf r})$ at the boundary of each NN
direction (i.e., the point A in Fig. \ref{fig:line}).    
The fourfold symmetry around the vortex core due to the
$d_{x^2-y^2}$-wave nature is seen in  
Figs. \ref{fig:h-high} (b) and (c), where $h({\bf r})$ is enhanced in 
the $0^\circ$ directions, and suppressed in the $45^\circ$ directions. 
In the $d45^\circ$ case in Fig. \ref{fig:h-high} (b), 
the $d_{x^2-y^2}$-wave nature enhances $h({\bf r})$ at the minimum point in  
the NNN direction, and suppresses it at the saddle point in the NN direction. 
So, variation of $h({\bf r})$ along the boundary line AC is small. 
On the other hand, for the $d0^\circ$ case in  
Fig. \ref{fig:h-high} (c), $h({\bf r})$ is enhanced at the saddle point
and suppressed at the minimum point due to the $d_{x^2-y^2}$-wave
nature. 
So, the difference between the field at the saddle point A and 
the field at the minimum point C is large. 
Since the $h({\bf r})$ distribution is observed in  
SANS and $\mu$SR experiments, we discuss the related quantities in detail. 

\begin{figure}
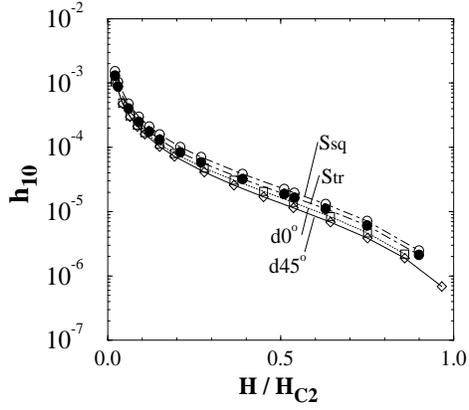

\caption{Field dependence of the dominant form factor $h_{1,0}$ 
for the $s$-wave ($\circ$), the $d45^\circ$ ($\diamond$) 
and the $d0^\circ$ ($\Box$) cases in the square lattice. 
The triangular lattice case of the $s$-wave pairing is also
plotted ($\bullet$).
}
\label{fig:h10}
\end{figure}

\begin{figure}
\leavevmode
\epsfxsize=8.0cm
\vspace{0.2cm}
\epsfbox{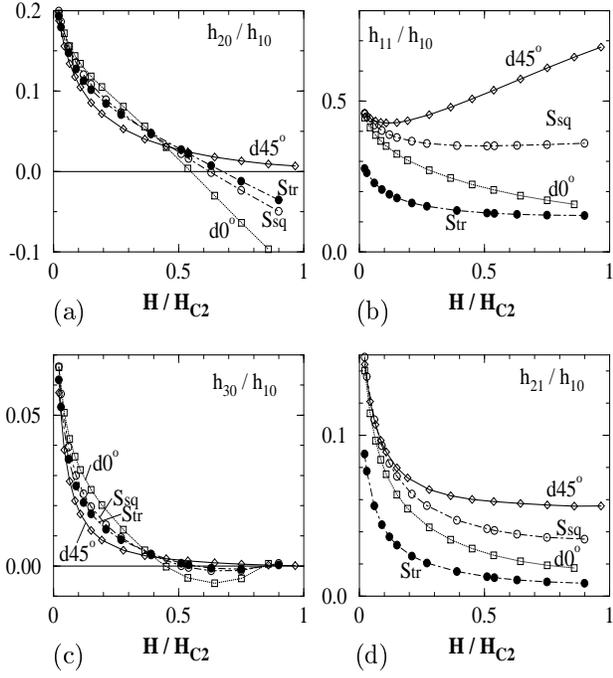}
\caption{Field dependence of higher order form factors. 
We plot $h_{2,0}/h_{1,0}$ (a), $h_{1,1}/h_{1,0}$ (b), 
$h_{3,0}/h_{1,0}$ (c) and  $h_{2,1}/h_{1,0}$ (d) 
for the $s$-wave ($\circ$), the $d45^\circ$ ($\diamond$) 
and the $d0^\circ$ ($\Box$) cases in the square lattice. 
The triangular lattice case of the $s$-wave pairing is also
plotted for reference ($\bullet$).
}
\label{fig:hmn}
\end{figure}

First, we investigate the form factor of $h({\bf r})$. 
It is measured by the SANS experiment.
The form factor $h_{m,n}$ ($m$ and $n$ are integers) is the Fourier
component of $h({\bf r})$ defined as 
\begin{equation}
h({\bf r})=H \sum_{m,n} h_{m,n} \exp(i {\bf g}_{m,n}\cdot {\bf r}) 
\label{eq:form-factor}
\end{equation}
where the reciprocal lattice vector ${\bf g}_{m,n}=-n{\bf k}_1 +m {\bf k}_2$, 
where ${\bf k}_1=2 \pi {\bf r}_2 \times {\bf r}_3 /S$, 
and ${\bf k}_2=2 \pi {\bf r}_3 \times {\bf r}_1 /S $ 
with ${\bf r}_3=(0,0,1)$ 
and $S={\bf r}_1 \cdot ({\bf r}_2 \times {\bf r}_3 )$ . 
The field dependence of the dominant form factor $h_{1,0}$ is plotted in
Fig. \ref{fig:h10} for the $s$- and $d_{x^2-y^2}$-wave pairings. 
The two pairing cases show a similar dependence on the applied
field. 
So, it is difficult to detect the $d_{x^2-y^2}$-wave nature in
$h_{1,0}$. 
However, we can detect it in the higher order components such as 
$h_{2,0}$ and $h_{1,1}$. 
The field dependence of $h_{2,0}/h_{1,0}$, $h_{1,1}/h_{1,0}$,
$h_{3,0}/h_{1,0}$ and $h_{2,1}/h_{1,0}$ is plotted in
Fig. \ref{fig:hmn}. 
The effect of the $d_{x^2-y^2}$-wave nature appears depending on the
orientation of the vortex lattice. 
As shown in Fig. \ref{fig:hmn} (a), with increasing field,
$h_{2,0}/h_{1,0}$ of the $s$-wave pairing decreases and becomes
negative. 
In the stable $d45^\circ$ case, $h_{2,0}/h_{1,0}$ remains positive. 
However, in the unstable $d0^\circ$ case, $h_{2,0}/h_{1.0}$
decreases more rapidly than that of the $s$-wave pairing. 
As shown in Fig. \ref{fig:hmn} (b), $h_{1,1}/h_{1,0}$ remains almost
constant at high field in the $s$-wave pairing. 
In the $d45^\circ$ case, $h_{1,1}/h_{1,0}$ increases 
as $H_{c2}$ is approached. 
But, in the $d0^\circ$ case, it decreases. 
For $\theta_0=0^\circ$ and for $\theta_0=45^\circ$, 
the $d_{x^2-y^2}$-wave effect  shifts the form factors 
in the opposite direction. 
The $d_{x^2-y^2}$-wave nature also appears in $h_{3,0}/h_{1,0}$
[Fig. \ref{fig:hmn} (c)] and $h_{2,1}/h_{1,0}$ [Fig. \ref{fig:hmn}
(d)]. 
We expect that these $d_{x^2-y^2}$-wave natures can be detected by 
the SANS experiment. 
The form factors are also presented for the triangular vortex lattice of the
$s$-wave pairing in Figs. \ref{fig:h10} and \ref{fig:hmn}. 
Their field dependence is qualitatively the same as in the square 
lattice case of the $s$-wave pairing. 

\begin{figure}
\leavevmode
\epsfxsize=6.0cm
\epsfbox{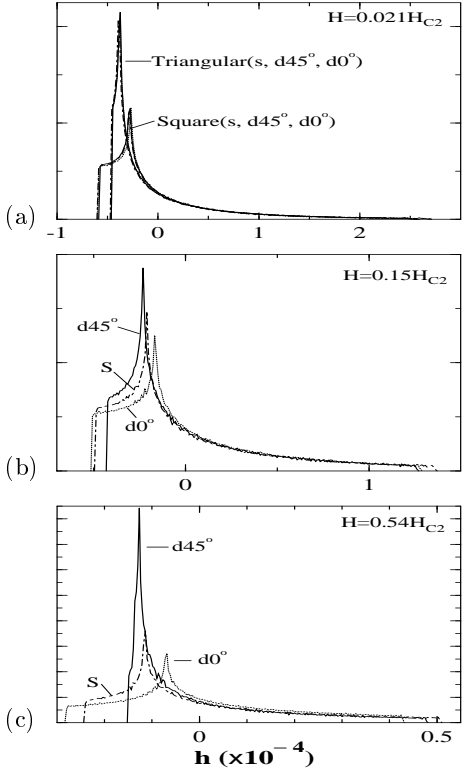}
\vspace{0.2cm}
\caption{Magnetic field distribution function $P(h)$ of the square
 vortex lattice case at $H/H_{c2}$=0.021 (a), 0.15 (b) and 0.54 (c) 
for the $s$-wave, the $d45^\circ$ and the $d0^\circ$ cases. 
$h$ is normalized by $\phi_0/\xi_0^2$. 
In (a), the lines for the triangular vortex lattice are also plotted. 
}
\label{fig:Ph}
\end{figure}

\begin{figure}
\leavevmode
\epsfxsize=6.0cm
\epsfbox{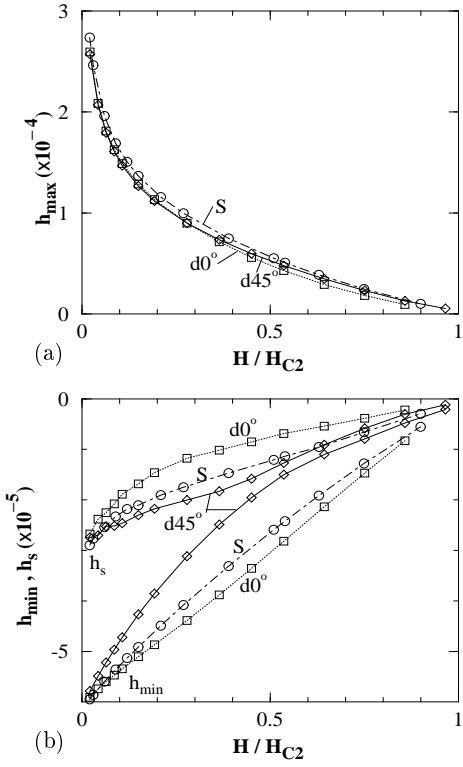}
\vspace{0.2cm}
\caption{Field dependence of $h_{\rm max}$  (a), $h_s$ and $h_{\rm min}$ (b) 
for the $s$-wave ($\circ$), the $d45^\circ$ ($\diamond$) and the
$d0^\circ$ ($\Box$) cases in the square lattice.
$h$ is normalized by $\phi_0/\xi_0^2$. 
}
\label{fig:hs}
\end{figure}

Next, we investigate the magnetic field distribution function defined as 
\begin{equation}
P(h)=\langle \delta(h-h({\bf r}))\rangle. 
\end{equation}
It corresponds to the resonance line shape in $\mu$SR or NMR
experiments. 
In Fig. \ref{fig:Ph}, we plot $P(h)$ for $H/H_{c2}$=0.021(a), 0.15(b) 
and 0.54(c) in the $s$-wave pairing case and in the two orientation 
cases of the $d_{x^2-y^2}$-wave pairing. 
We denote the maximum (minimum) of $h({\bf r})$ as $h_{\rm max}$ 
($h_{\rm min}$). 
Reflecting the saddle point in $h({\bf r})$, $P(h)$ has a logarithmic
singularity  at $h_s$. 
In the low field case shown in Fig. \ref{fig:Ph} (a), $P(h)$ is almost the same
shape for both the $s$- and $d_{x^2-y^2}$-wave pairings. 
For comparison, we also present the line shape of the triangular lattice 
case in Fig.  \ref{fig:Ph} (a), which also does not show any difference 
between the $s$- and $d_{x^2-y^2}$-wave pairings.
As shown in Fig. \ref{fig:lowH} (c), the $d_{x^2-y^2}$-wave nature
scarcely appears in $h({\bf r})$ at a low field. 
With increasing field, the $d_{x^2-y^2}$-wave nature gradually appears in 
$P(h)$ as shown in Figs. \ref{fig:Ph} (b) and (c). 
For the stable $d45^\circ$ case,  $h_{\rm min}$ increases and 
$h_s$  decreases compared with the $s$-wave pairing case, 
as shown in Fig. \ref{fig:h-high} (b). 
For the unstable $d0^\circ$ case, $h_{\rm min}$ decreases and 
$h_s$ increases. 
These changes of $h_{\rm min}$ and $h_s$ can be seen in the resonance line
shape in Figs. \ref{fig:Ph} (b) and (c). 
We note that the line shape of the $d45^\circ$ case in Fig. \ref{fig:Ph} 
(c) resembles that of the triangular lattice (see Fig. \ref{fig:Ph} (a)). 
So, in the $d_{x^2-y^2}$-wave pairing, 
it is risky to determine the vortex lattice shape by the
conventional method based on the distance between $h_s$ and 
$h_{\rm min}$, i.e., $(h_s-h_{\rm min})/(h_{\rm max}-h_{\rm min})$. 
The field dependence of $h_{\rm max }$, $h_s$ and $h_{\rm min}$ is
presented in Fig. \ref{fig:hs}. 
There, $h_{\rm max}$ shows a similar dependence for both the $s$- and 
$d_{x^2-y^2}$-wave pairings. 
We also calculate the field dependence of $h_{\rm max }$ in the
triangular vortex lattice for the $s$-wave and $d_{x^2-y^2}$-wave pairings. 
Their dependence is  almost the same as that of the square lattice case. 
So, $h_{\rm max }$ is not much affected by the vortex lattice configuration 
and the pairing symmetry. 
The $d_{x^2-y^2}$-wave nature affects $h_{\rm min}$ and $h_s$. 
For the stable $d45^\circ$ case, 
$h_{\rm min}$ approaches $h_s$ with increasing field, and $h_{\rm min}
\sim h_s$ for $H/H_{c2} > 0.5$. 
We hope that this characteristic of the resonance line shape $P(h)$ will 
be examined by the $\mu$SR or NMR experiments. 

\begin{figure}
\leavevmode
\epsfxsize=8.0cm
\epsfbox{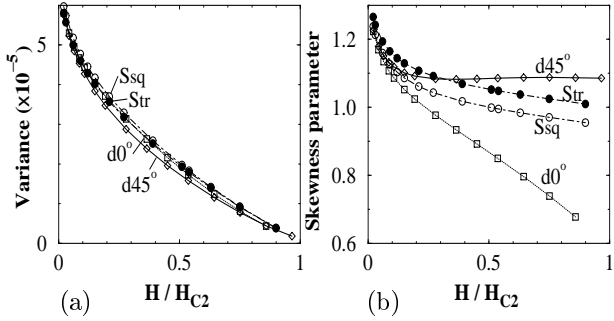}
\vspace{0.2cm}
\caption{Field dependence of variance $\sigma$ (a) and  skewness parameter 
$\alpha$ (b) for the $s$-wave ($\circ$), the $d45^\circ$ ($\diamond$) 
and the $d0^\circ$ ($\Box$) cases in the square lattice. 
The triangular lattice case of the $s$-wave pairing is also
plotted ($\bullet$).
}
\label{fig:variance}
\end{figure}

To help in analyzing the shape of the distribution function $P(h)$, 
we plot the field dependence of the variance 
$\sigma =\langle h({\bf r})^2 \rangle^{1/2}$ and of the skewness 
parameter $\alpha=\langle h({\bf r})^3 \rangle^{1/3}/\sigma$ 
in Fig. \ref{fig:variance}.
These quantities are analyzed by the $\mu$SR 
experiment.\cite{Yaouanc,Reotier,Lee,Aegerter} 
As for the variance, both the $s$- and $d_{x^2-y^2}$-wave pairings show a 
similar dependence on $H$. 
But, the difference between the $s$- and $d_{x^2-y^2}$-wave pairings 
clearly appears in the skewness parameter.

\section{Vortex structure in the triangular lattice case}
\label{sec:triangular}

\begin{figure*}
\leavevmode
\epsfxsize=12.0cm
\epsfbox{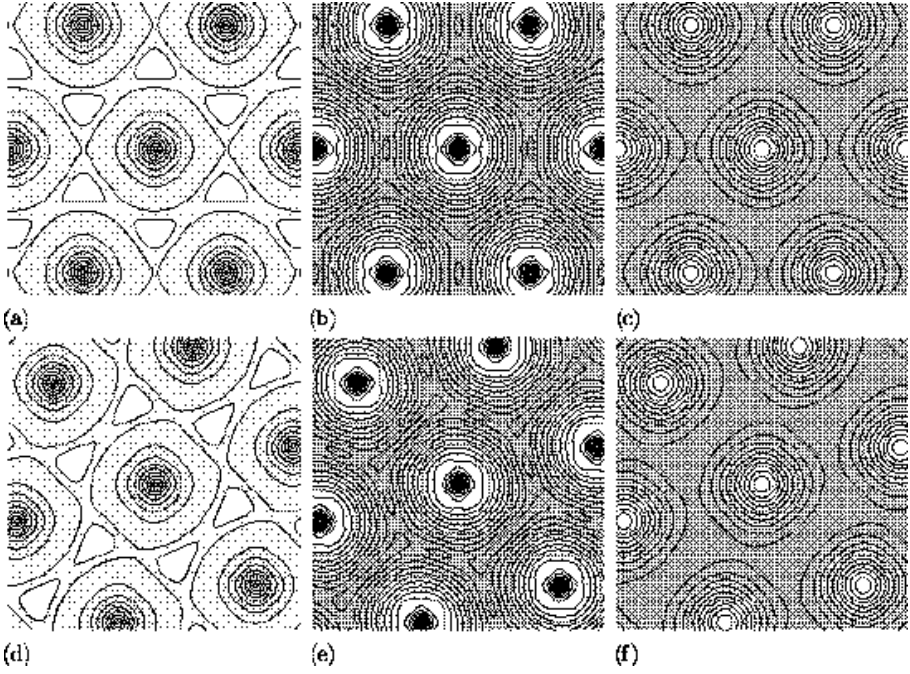}
\vspace{0.2cm}
\caption{Contour plot of the vortex structure 
in the triangular vortex lattice for the $d_{x^2-y^2}$-wave pairing. 
As for the orientation, $\theta_0=0^\circ$ [(a)-(c)] 
and $45^\circ$ [(d)-(f)]. 
The amplitude $|\Delta({\bf r})|$ [(a) and (d)], 
the current $|{\bf J}({\bf r})|$ [(b) and (e)], 
and the internal field $h({\bf r})$ [(c) and (f)] 
at $H/H_{c2}=0.15$ are presented. 
The $a$ axis (i.e., $0^\circ$ direction) is horizontal. 
To show the distortion of the contour lines clearly in (a) and (d), 
we also present the contour line for 
$|\Delta({\bf r})|/\Delta_0=0.89$ in addition to the 
lines of 0.1, $\cdots$, 0.9.  
}
\label{fig:triangular}
\end{figure*}

To study the vortex structure at low field and the effect of the vortex 
lattice shape, we also consider the triangular vortex lattice case. 
In Fig. \ref{fig:triangular}, we show the vortex structure at
$H/H_{c2}=0.15$ for the $d_{x^2-y^2}$-wave pairing. 
The contour lines of $|\Delta({\bf r})|$, $|{\bf J}({\bf r})|$ 
and $h({\bf r})$ are presented for the two orientation cases  
$\theta_0=45^\circ$ and $0^\circ$. 
The contour lines of $|\Delta({\bf r})|$ show a fourfold-symmetric shape, 
$|{\bf J}({\bf r})|$ has four peaks in the $45^\circ$ directions, 
and $h({\bf r})$ extends in the $0^\circ$ direction around each vortex core. 
These features show the $d_{x^2-y^2}$-wave nature 
presented in the previous section. 
At a low field such as $H/H_{c2}=0.021$, the fourfold-symmetric structure 
of the $d_{x^2-y^2}$-wave nature is restricted to the small region 
of the vortex core as in the case shown in Fig. \ref{fig:lowH}. 
These core structures are not affected by the shape and the orientation 
of the vortex lattice at low field. 
But, with increasing field, the fourfold-symmetric core structure gradually 
feels the effect of the hexagonal Wigner-Seitz cell of the vortex lattice. 
By the effect of the fourfold-symmetric structure around the core
region, the six NN directions of the triangular lattice are not equivalent 
in the  $d_{x^2-y^2}$-wave pairing. 
In Fig. \ref{fig:triangular} (a) for $\theta_0=0^\circ$, the suppression 
of $|\Delta({\bf r})|$ in the NN directions due to the vortex lattice
effect is enhanced in the $a$ axis direction by the $d_{x^2-y^2}$-wave
nature. 
So, the structure of the two NN directions along the $a$ axis is not
equivalent to that of the other four NN directions. 
The suppression due to the $d_{x^2-y^2}$-wave nature in the $b$ axis
direction is weakened by the vortex lattice effect because it is in the NNN 
direction. 
The vortex structure now shows twofold symmetry, which becomes more
prominent at higher field. 
In Fig. \ref{fig:triangular} (d) for $\theta_0=45^\circ$, 
the suppression in the NN direction is weakened in the $45^\circ$
direction, compared with the other four NN directions. 
 
\begin{figure}
\leavevmode
\epsfxsize=6.0cm
\epsfbox{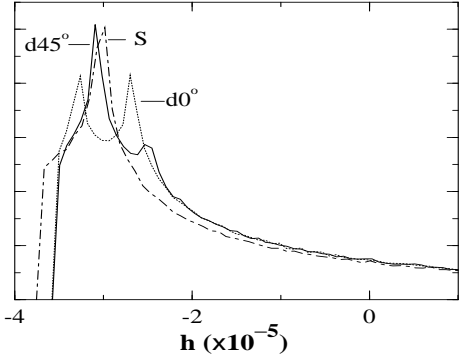}
\vspace{0.2cm}
\caption{
Magnetic field distribution function $P(h)$ of the triangular 
 vortex lattice case at $H/H_{c2}$= 0.15 
for the $s$-wave, $d45^\circ$ and $d0^\circ$ cases.
}
\label{fig:Ph-tri}
\end{figure}

The structure of these two NN directions can also be seen in the
current $|{\bf J}({\bf r})|$ [Figs. \ref{fig:triangular} (b) and (e)] 
and the internal field $h({\bf r})$ [Figs. \ref{fig:triangular} (c) and
(f)].  
In Fig. \ref{fig:triangular} (c) for $\theta_0=0^\circ$, $h({\bf r})$ at 
the two saddle points in the $a$ axis NN direction is larger than that of
the saddle points in the other four  NN directions. 
In Fig. \ref{fig:triangular} (f) for $\theta_0=45^\circ$, $h({\bf r})$ at 
the two saddle points in the $45^\circ$ direction 
becomes small compared to the other four  NN directions. 
The resonance line shape $P(h)$ in the triangular lattice case is presented 
in Fig. \ref{fig:Ph-tri}, where the line shape around the peak is focused. 
Reflecting the two saddle points of $h({\bf r})$ in
Figs. \ref{fig:triangular} (c) and (f), the line shape of $P(h)$ has
two logarithmic peaks in the $d_{x^2-y^2}$-wave pairing. 
The two peak structure was suggested for 
the oblique lattice state  between the square lattice and
triangular lattice, based on the GL theory or the London
theory.\cite{Brandt,IchiokaGL,Berlinsky}  
But, it occurs even in the triangular lattice for the $d_{x^2-y^2}$-wave
pairing by the influence of the fourfold-symmetric vortex core structure. 

Relaxing the difference between these two  NN directions, the shape of 
the vortex lattice may deviate from the $60^\circ$ triangular lattice. 
For $\theta_0=0^\circ$ in Fig. \ref{fig:triangular} (a), the
intervortex distance in the $a$ axis NN directions may increase compared
with the other NN directions. 
For $\theta_0=0^\circ$ in Fig. \ref{fig:triangular} (d), the
intervortex distance in the $45^\circ$ directions  may
decrease  compared with the other NN directions. 
This deformation may relax the difference between the two  NN
directions. 
This is the origin of the distortion from the conventional $60^\circ$
triangular lattice. 
The oblique lattice is realized in the intermediate field
region between the square lattice and the $60^\circ$ triangular lattice.
In the oblique lattice, the double-peaked structure of $P(h)$ shown in 
Fig. \ref{fig:Ph-tri} may be smeared, since the difference of the two
saddle points is weakened by the distortion from the $60^\circ$
triangular lattice. 

We also calculate the field dependence of the form factor $h_{m,n}$ in
the triangular lattice case. 
It shows a similar dependence to the square lattice case presented in 
Figs. \ref{fig:h10} and \ref{fig:hmn}. 
The $d_{x^2-y^2}$-wave nature does not much affect $h_{1,0}$, but appears
in the higher order factors such as $h_{1,1}$ and $h_{2,0}$.

\section{Density of states}
\label{sec:dos}

\begin{figure*}
\leavevmode
\epsfxsize=16.0cm
\vspace{0.2cm}
\epsfbox{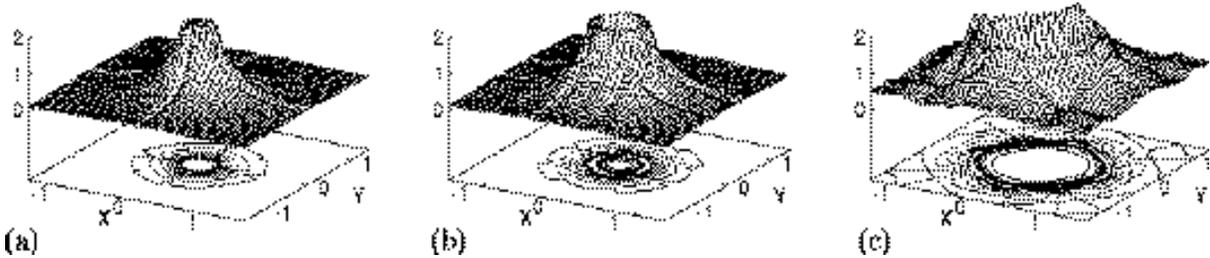}
\caption{
Spatial variation of the LDOS, $N(E,{\bf r})/N_0$, for the s-wave 
pairing at $H/H_{c2}$= 0.15.  
$E/\Delta_0=0$ (a), 0.4 (b), and 0.8 (c). 
The region of the Wigner-Seitz cell shown in Fig. \protect\ref{fig:line} is 
 presented. 
In these figures, the peak is truncated at $N(E,{\bf r})/N_0=2$. 
}
\label{fig:ldos-s}
\end{figure*}

The LDOS $N(E,{\bf r})$ is calculated by Eq. (\ref{eq:ldos}) in the
square vortex lattice case for the $s$-wave and the $d45^\circ$ cases. 
Typically, we choose $\eta=0.03$. 
First, we discuss the LDOS for the $s$-wave pairing. 
In the calculation for a single vortex, the LDOS for low energy
excitations is mainly distributed in a circle around the vortex 
center.\cite{Ullah,Schopohl} 
The radius of the circle increases if $E$ is raised. 
The LDOS of the vortex lattice case is presented in
Fig. \ref{fig:ldos-s} for the $s$-wave pairing at $H/H_{c2}=0.15$. 
The area presented is the Wigner-Seitz cell of the square vortex
lattice, i.e., within the dotted line shown in Fig. \ref{fig:line}. 
The LDOS for low energy excitations is distributed in 
a circle as in the single vortex case. 
Further, by the vortex lattice effect (i.e., the quasiparticle transfer
between vortices), $N(E=0,{\bf r})$ is suppressed on the line connecting 
two NN vortex centers, as shown in Fig. \ref{fig:ldos-s} (a). 
A small suppression also occurs along the line between two NNN vortex
centers. 
For a finite $E$ below $\Delta_0$, the suppression occurs in the NN
directions along the tangent lines of the circle of the dominant 
LDOS distribution, as shown in Fig. \ref{fig:ldos-s} (b) and (c).  
The suppression is scarcely seen at low field $(H < 0.1 H_{c2})$. 
There, the LDOS distribution shows an almost circular symmetry 
around each vortex. 
At high field, however, the suppression becomes significant and 
the LDOS distribution deviates from the circular 
symmetry and shows fourfold-(sixfold-)symmetric structure in the
square (triangular) lattice case, reflecting the vortex lattice 
effect.\cite{IchiokaS}
At high field, the LDOS has a finite distribution all over the unit cell 
in addition to the ridge structure, while at low
field it is distributed in a restricted region along the circular ridge.

\begin{figure*}
\leavevmode
\epsfxsize=16.0cm
\vspace{0.2cm}
\epsfbox{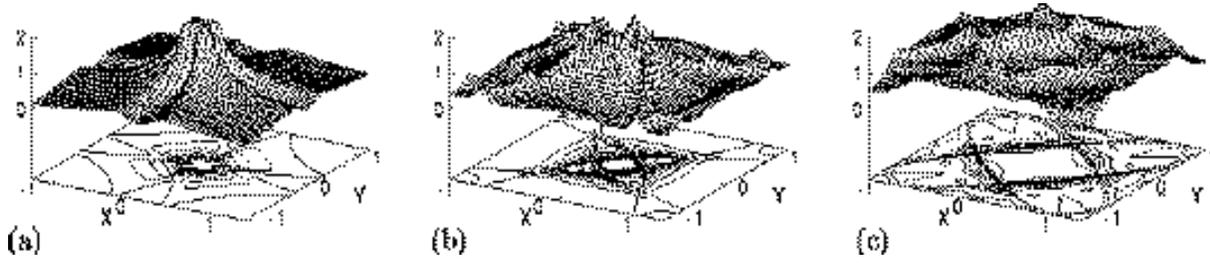}
\caption{
The same as Fig. \protect\ref{fig:ldos-s}, but for the 
$d_{x^2-y^2}$-wave pairing instead of the $s$-wave pairing, 
and $E/\Delta_0=0$ (a), 0.5 (b), 1 (c). 
}
\label{fig:ldos-d}
\end{figure*}

Next, we discuss the $d_{x^2-y^2}$-wave pairing case. 
The single vortex calculation shows that the LDOS for low energy
excitations around a vortex has prominent tails extending in the
$45^\circ$ directions [see Fig. 8 (d) of Ref. \onlinecite{IchiokaD}], 
reflecting the node structure of the energy gap in 
the $d_{x^2-y^2}$-wave pairing. 
Since the low energy quasiparticle distribution around a vortex extends
toward infinite points in the $45^\circ$ directions, this low
energy state is not a bound state in the exact meaning.\cite{FranzBdG} 
The LDOS of the vortex lattice case is presented in Fig. \ref{fig:ldos-d} 
at $H/H_{c2}=0.15$ for the $d45^\circ$ case.  
In this configuration, the $45^\circ$ directions are the NN vortex
directions, which are on the $x$ axis and $y$ axis in the figure. 
In $N(E=0,{\bf r})$ in Fig. \ref{fig:ldos-d} (a), 
the tail structure in the $45^\circ$ directions
in the single vortex case is modified due to the vortex lattice effect. 
Each tail of the zero energy LDOS in the $45^\circ$ directions splits into
two ridges because of the suppression by the vortex lattice effect in
the NN vortex directions. 
So, the tails extend in rather different directions from the
$45^\circ$ directions. 
In the $d_{x^2-y^2}$-wave pairing, the suppression of the vortex lattice 
effect occurs from a low field such as $H/H_{c2}=0.02$. 
It means that the quasiparticle transfer between vortices is large due to 
the tail structure of the LDOS in the $d_{x^2-y^2}$-wave pairing,
compared with the $s$-wave pairing case.
With increasing field, the vortex lattice effect increases and 
the splitting of the tail structure in the $45^\circ$ directions 
becomes pronounced. 

For a finite $E$ below the energy gap $\sqrt{2}\Delta_0$, 
$N(E,{\bf r})$ shows a ``\#''-shape 
ridge structure in the $d_{x^2-y^2}$-wave pairing, as shown in 
Fig. \ref{fig:ldos-d} (b) and (c). 
This is consistent with the calculation for the 
single vortex case.\cite{IchiokaD,Schopohl} 
The ridge structure is not much affected by the vortex lattice effect 
except when $E \sim 0$.  
The reason is as follows. 
The quasiparticle flow is along the ridges 
on the four open trajectories.\cite{IchiokaD} 
When the flow direction $\theta$ of the quasiparticles on the trajectories
approaches the $45^\circ$ directions, it leaves the vortex
center.  
There, the quasiparticles with a finite $E$ become a scattering state, 
since $\sqrt{2}\Delta_0 |\phi(\theta)|< E$ is satisfied. 
When the quasiparticle flow is toward the NN vortex directions, the vortex
lattice effect occurs. 
However, since the NN directions are now the $45^\circ$ directions, 
the quasiparticles flowing toward the NN vortex directions on the
trajectories are in a scattering state. 
A scattering state is widely distributed and does not contribute to the
ridge structure. 

As a result, the local spectrum $N(E,{\bf r})$ is given as follows in
the vortex lattice state. 
There are some peaks below the energy gap as in the single vortex case, 
reflecting the trajectory structure.\cite{IchiokaD} 
Extra small peaks may also appear by the trajectories extending
from neighboring vortices. 
And the low energy state for $E \sim 0$ is suppressed 
due to the vortex lattice effect 
on the line connecting two NN (or NNN) vortex centers.

We have considered the case of the square lattice with
$\theta_0=45^\circ$. 
From that information, 
we can estimate the spatial variation of $N(E=0,{\bf r})$ in the
other vortex lattice configuration. 
First, we consider the contribution on $N(E=0,{\bf r})$ from the vortex
core and the four tails extending in the $45^\circ$ 
directions from each vortex center. 
The suppression along the line connecting NN vortices is also taken into
account. 
There is also a small suppression in the NNN vortex directions. 
The suppression is larger at higher field. 
Then, we obtain the $N(E=0,{\bf r})$ structure in the other vortex lattice
configuration. 

The relation $N(E=0)\propto\sqrt{H}$ is sometimes considered for the
$d_{x^2-y^2}$-wave pairing. 
The relation is based on the single vortex calculation,\cite{Volovik} 
which yields a structure with four tails extending 
in the $45^\circ$ directions.  
These result from the node structure of the energy gap. 
However, since the vortex lattice effect smears the tail structure, the
relation $N(E=0)\propto\sqrt{H}$ may be affected. 
We examined it in Ref. \onlinecite{IchiokaL} and obtained $N(E=0)\propto
H^{0.41}$. 
The exponent becomes smaller than $1/2$. 
The relation $N(E=0)\propto H$ is sometimes considered for the $s$-wave
pairing. In this case, $N(E=0)$ is proportional to the
vortex density since the low energy excitation consists of a bound
state around each vortex core. 
We also examined this case in Ref. \onlinecite{IchiokaL} and obtained 
$N(E=0)\propto H^{0.67}$. 
The deviation from the $H$-linear relation results from the field
dependence of the vortex core radius which we presented in 
Fig. \ref{fig:radius}. 

\begin{figure}
\leavevmode
\epsfxsize=8.0cm
\epsfbox{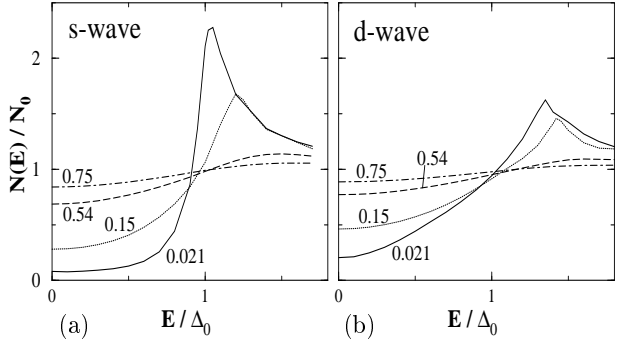}
\vspace{0.2cm}
\caption{
Spectrum of the spatially averaged DOS, $N(E)/N_0$,  
at $H/H_{c2}$=0.021, 0.15, 0.54 and 0.75 for the $s$-wave (a) 
and the $d_{x^2-y^2}$-wave (b) pairings. 
}
\label{fig:dos}
\end{figure}

The spatially averaged DOS in Eq. (\ref{eq:dos}) is presented in
Fig. \ref{fig:dos} for the $s$-wave (a) and $d_{x^2-y^2}$-wave (b) pairings.  
We note that the energy gap in the $d_{x^2-y^2}$-wave case is
$\sqrt{2}\Delta_0$ in Fig. \ref{fig:dos} (b). 
The energy gap structure at zero field 
is gradually buried by the low energy excitation 
of the vortex as the field increases. 
While the local spectrum $N(E,{\bf r})$ has several peaks 
depending on the pairing symmetry and the position,\cite{IchiokaD}  
there is no peak  in $N(E)$ below the energy gap. 
The peak structure is smeared when the spatial average is taken.

\section{Induced order parameter of other symmetry}
\label{sec:induced}

So far, we have considered  the case where we can neglect the induced order 
parameter of other symmetry, i.e., the pure $d_{x^2-y^2}$-wave case. 
When there is $s$-wave ($d_{xy}$-wave) interaction in addition to 
the dominant $d_{x^2-y^2}$-wave pairing interaction, 
the $s$-wave ($d_{xy}$-wave) component of the pair potential is 
induced at the place where the dominant $d_{x^2-y^2}$-wave pair potential 
spatially varies, such as a vortex or an interface. 
The induced $s$-wave component was mainly studied by the two-component
GL theory, where the coupling of the $s$-wave and the 
$d_{x^2-y^2}$-wave components in the gradient term plays an important
role.\cite{Berlinsky,Ren}
The induced $d_{xy}$-wave component can be discussed by extending the two
component GL theory by including the higher order gradient
coupling.\cite{IchiokaInd} 
The quasi-classical calculation for the induced $s$-wave and
$d_{xy}$-wave components was performed for the single vortex
case.\cite{IchiokaInd} 
The induced component around the vortex was also reported in a study
of the Bogoliubov-de Gennes theory for several
models.\cite{FranzBdG,Soininen,Wang,Himeda} 
There, the amplitude of the induced component depends on the
interaction parameter and the carrier density. 

In this section, we study the spatial structure of the induced $s$- and
$d_{xy}$-wave components in the vortex lattice of  $d_{x^2-y^2}$-wave
superconductors, based on the quasi-classical theory. 
These induced components can be calculated using Eq. (\ref{eq:self-d}), 
by setting  
\begin{eqnarray} 
\Delta(\theta,{\bf r})
&=&
\Delta_s({\bf r})\phi_s(\theta)
+\Delta_{x^2-y^2}({\bf r})\phi_{x^2-y^2}(\theta) 
\nonumber \\ &&
+\Delta_{xy}({\bf r})\phi_{xy}(\theta) , 
\label{eq:delta-mix}
\end{eqnarray}
\begin{eqnarray} 
V(\theta',\theta)
&=&
\bar{V}_s \phi_s(\theta') \phi_s(\theta)
+\bar{V}_{x^2-y^2} \phi_{x^2-y^2}(\theta') \phi_{x^2-y^2}(\theta) 
\nonumber \\ &&
+\bar{V}_{xy} \phi_{xy}(\theta') \phi_{xy}(\theta) 
\label{eqv-mix}
\end{eqnarray}
where $\phi_s(\theta)=1$, 
$\phi_{x^2-y^2}(\theta)=\sqrt{2}\cos 2(\theta -\theta_0)$ and 
$\phi_{xy}(\theta)=\sqrt{2}\sin 2(\theta -\theta_0)$. 
Note that the induced components $\Delta_s({\bf r})$ and 
$\Delta_{xy}({\bf r})$ are, respectively, proportional to 
$\bar{V}_s$ and $\bar{V}_{xy}$. 
Here, we consider the case 
$|\bar{V}_s|,|\bar{V}_{xy}| \ll |\bar{V}_{x^2-y^2}|$. 
In this case, the contributions of $\Delta_s({\bf r})$ and 
$\Delta_{xy}({\bf r})$ are small and can be neglected compared to 
the dominant $\Delta_{x^2-y^2}({\bf r})$ contribution 
in the calculation of Eqs. (\ref{eq:eil-1})-(\ref{eq:eil-3}).  

\begin{figure}
\leavevmode
\epsfxsize=8.0cm
\epsfbox{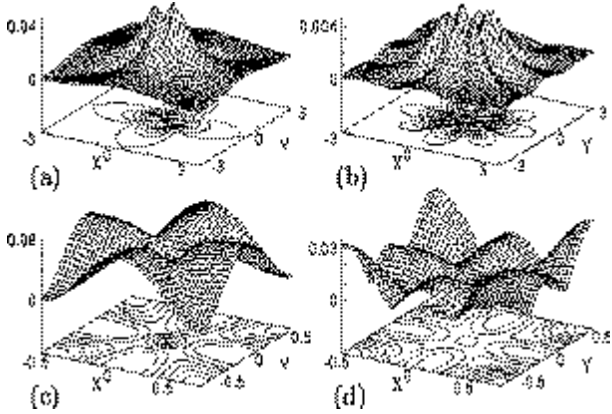}
\vspace{0.2cm}
\caption{
Spatial variation of the induced $s$-wave and $d_{xy}$-wave components  
at low field $H/H_{c2}=0.021$ [(a) and (b)] and high field 
$H/H_{c2}=0.54$ [(c) and (d)]. 
The amplitude $|\Delta_s({\bf r})|/|\bar{V}_s/\bar{V}_{x^2-y^2}|$ 
[(a) and (c)] and $|\Delta_{xy}({\bf r})|/|\bar{V}_{xy}/\bar{V}_{x^2-y^2}|$ 
[(b) and (d)] are presented in the Wigner-Seitz cell region shown in 
Fig. \protect\ref{fig:line}. 
}
\label{fig:ind-amp}
\end{figure}

Figure \ref{fig:ind-amp} shows the spatial variation of the amplitudes  
$|\Delta_s({\bf r})|$ and $|\Delta_{xy}({\bf r})|$ in the $d45^\circ$ case
at low field $H/H_{c2}=0.021$ and high field $H/H_{c2}=0.54$. 
The presented area is the Wigner-Seitz cell of the square vortex
lattice, i.e., within the dotted line in Fig. \ref{fig:line}. 
At low field, the contour line of $|\Delta_s({\bf r})|$ is a four-lobed  
shape, as shown in Fig.  \ref{fig:ind-amp} (a). 
There,  $|\Delta_s({\bf r})|$ has four peaks around the vortex core
in the $45^\circ$ directions (which correspond to the $x$ axis and $y$ axis
in this figure). 
The contour line of $|\Delta_{xy}({\bf r})|$  is eight-lobed shape shape, as
shown in Fig.  \ref{fig:ind-amp} (b). 
There are eight peaks of $|\Delta_{xy}({\bf r})|$ around the vortex core
in the $0^\circ$ and $45^\circ$ directions. 
These structures of $|\Delta_s({\bf r})|$ and $|\Delta_{xy}({\bf r})|$ at 
low field are consistent with those found in a single vortex.\cite{IchiokaInd} 
With increasing field, since the size of  the unit cell decreases, the
structures of $|\Delta_s({\bf r})|$ and $|\Delta_{xy}({\bf r})|$
gradually change to those shown in Figs. \ref{fig:ind-amp} (c) and (d),
respectively. 
These structures have extra zero points on the boundary in addition to
the vortex center. 

\begin{figure}
\leavevmode
\epsfxsize=8.0cm
\epsfbox{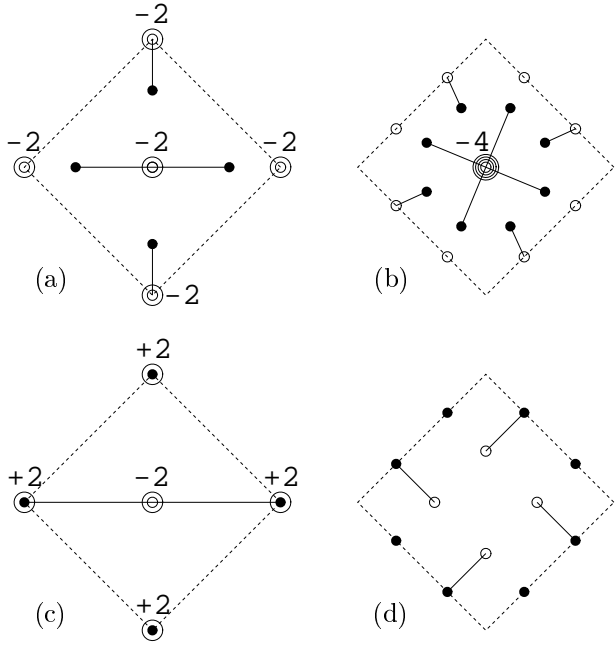}
\vspace{0.2cm}
\caption{
Spatial structure of the relative phase for the induced $s$-wave 
component [(a)  and (c)] and the $d_{xy}$-wave component [(b) and (d)]  
at low field [(a) and (b)] and high field [(c) and (d)]. 
The position of the singularity and its winding number are presented 
schematically for ${\rm arg}(\Delta_s({\bf r})/\Delta_{x^2-y^2}({\bf r}))$ 
and ${\rm arg}(\Delta_{xy}({\bf r})/\Delta_{x^2-y^2}({\bf r}))$. 
$\bullet$ and $\circ$ mean the singularities with 
winding numbers $+1$ and $-1$, respectively. 
The solid lines show the cut of the phase, where the phase jumps from 
$-\pi$ to $\pi$. 
The dotted line shows the Wigner-Seitz cell shown in  
Fig. \protect\ref{fig:line}. 
}
\label{fig:ind-phase}
\end{figure}

To understand the structures of $\Delta_s({\bf r})$ and
$\Delta_{xy}({\bf r})$, we present the structure of the phase 
schematically for low and high field in Fig. \ref{fig:ind-phase}. 
There, the position of the singularity and its winding number are
presented for the relative phase to the dominant $\Delta_{x^2-y^2}({\bf r})$, 
i.e., ${\rm arg}(\Delta_s({\bf r})/\Delta_{x^2-y^2}({\bf r}))$ and 
${\rm arg}(\Delta_{xy}({\bf r})/\Delta_{x^2-y^2}({\bf r}))$. 
Since $\Delta_{x^2-y^2}({\bf r})$ has the singularity of the winding
number $+1$ at the vortex center, the winding number of 
${\rm arg}(\Delta_s({\bf r}))$ itself is $-1$ at the vortex center 
[the case shown in Fig. \ref{fig:ind-phase} (a) and (c)]. 
And that of ${\rm arg}(\Delta_{xy}({\bf r}))$ is $-3$ at low field 
[Fig. \ref{fig:ind-phase} (b)] and $+1$ at high field 
[Fig. \ref{fig:ind-phase} (d)].  
The singularity corresponds to the zero point of the amplitudes  
$|\Delta_s({\bf r})|$ and $|\Delta_{xy}({\bf r})|$ in
Fig. \ref{fig:ind-amp}. 
Around the singularity of the winding number $n$, the amplitude
increases as $r^{|n|}$. 

At low field, as shown in Fig. \ref{fig:ind-phase} (a), the relative phase
of the $s$-wave component has a $-2$ singularity at the vortex center
and four $+1$ singularities in the $0^\circ$ directions. 
This is consistent with the single vortex case.\cite{IchiokaS}  
Since the winding number of the relative phase should total zero for each 
unit cell to satisfy the periodic boundary condition, 
there is a $-2$ singularity at each corner of the Wigner-Seitz cell. 
With increasing field, the $+1$ singularity approaches the boundary 
and the four $+1$ singularities combine with the $-2$ singularities at the
corners producing  $+2$ singularities, as shown in
Fig. \ref{fig:ind-phase} (c). 
In the triangular vortex lattice case, each $+2$ ($-2$) singularity at
the corner of the boundary splits into two $+1$ ($-1$) singularities 
which are located at the corner of the hexagonal Wigner-Seitz cell. 

On the other hand, 
${\rm arg}(\Delta_{xy}({\bf r})/\Delta_{x^2-y^2}({\bf r}))$ has a $-4$
singularity at the vortex center and eight $+1$ singularities 
around the vortex core at low field.  
This is consistent with the single vortex case.\cite{IchiokaInd} 
Further, to satisfy the periodic boundary condition, there are eight
$-1$ singularities on the boundary, as shown in Fig. \ref{fig:ind-phase}
(b). 
As the field increases, the $+1$ singularities approach the $-1$
singularities on the boundary, becoming  $+1$ singularities on the
boundary, as shown in Fig. \ref{fig:ind-phase} (d). 
The $-4$ singularity at the vortex center splits into four $-1$
singularities around the vortex core.

We also consider the $d0^\circ$ case, where the $0^\circ$ direction is
along the $x$ and $y$ axes. 
The four-peaked structure of $|\Delta_s({\bf r})|$ at low field rotates 
 $45^\circ$ from its position in the $d45^\circ$ case 
[Fig. \ref{fig:ind-amp} (a)]. 
The low field structure is fixed to the crystal coordinate. 
However, as the field increases, the structure of $\Delta_s({\bf r})$ reduces 
to that shown in Fig. \ref{fig:ind-amp} (c). 
The high field structure is independent of the crystal coordinate and
is fixed to the shape of the vortex lattice. 
On the other hand, the structure of $\Delta_{xy}({\bf r})$ is the same
as in the $d45^\circ$ case regardless of the field. 

Even in an $s$-wave superconductor, if the extra $d_{x^2-y^2}$-wave
($d_{xy}$-wave) component exists in the interaction,  
$\Delta_{x^2-y^2}({\bf r})$ ($\Delta_{xy}({\bf r})$) is induced around
the vortex. 
We also calculate its structure in the vortex configuration shown in 
Fig. \ref{fig:damp-high} (a). 
The induced $\Delta_{x^2-y^2}({\bf r})$ ($\Delta_{xy}({\bf r})$) has the 
same structure as $\Delta_s({\bf r})$ in the $d45^\circ$ 
($d0^\circ$) case. 

The induced component may be discussed qualitatively by the two
component GL theory.\cite{IchiokaGL,Berlinsky,Ren,IchiokaInd}
There, the vortex structure was studied in the single vortex case near 
$H_{c1}$ and in the vortex lattice state near $H_{c2}$. 
The results agree with ours. 
Near $H_{c2}$, the vortex structure is described by the magnetic Bloch
state function $\psi_n({\bf r}|{\bf r}_0)$, where $n$ represents the $n$-th
Landau level.\cite{IchiokaGL} 
As for the dominant $d_{x^2-y^2}$-wave pair potential, 
$\Delta_{x^2-y^2}({\bf r}) \sim \psi_0({\bf r}|{\bf r}_0)$ in the
leading order. 
For the induced $s$-wave component, 
$\Delta_s({\bf r})\sim \psi_2({\bf r}|{\bf r}_0)$ 
since it is induced by the gradient coupling 
$(\partial_x^2 - \partial_y^2) \Delta_{x^2-y^2}$ in the GL equation. 
For the induced $d_{xy}$-wave component, 
$\Delta_{xy}({\bf r})\sim \psi_4({\bf r}|{\bf r}_0)$ 
since it is induced by the gradient coupling 
$\partial_x \partial_y (\partial_x^2 - \partial_y^2)\Delta_{x^2-y^2}$ 
in the extended GL equation.\cite{IchiokaInd} 
In fact, at high field, $\Delta_s({\bf r})$ and $\Delta_{xy}({\bf r})$ 
as obtained by our calculation have the same structure as 
$\psi_2({\bf r}|{\bf r}_0)$ and  $\psi_4({\bf r}|{\bf r}_0)$,
respectively. 

\section{Summary and Discussions}
\label{sec:summary}

We have studied the vortex structure and its field dependence in the
$d_{x^2-y^2}$-wave and $s$-wave superconductors by the self-consistent
calculation of the quasi-classical Eilenberger theory for the vortex
lattice case. 
At a higher field, the square vortex lattice configuration 
with $\theta_0=45^\circ$ (i.e., the NN vortex is located in a $45^\circ$
direction from the $a$ axis) has lower free energy than the conventional 
$60^\circ$ triangular lattice for the $d_{x^2-y^2}$-wave pairing. 
Then, we investigated the field dependence of the vortex structure in the 
square lattice configuration. 
Due to the $d_{x^2-y^2}$-wave nature, the contour lines of the amplitude
$|\Delta({\bf r})|$ and the internal field $h({\bf r})$ show a 
fourfold-symmetric shape around the vortex core.  
The fourfold-symmetric structure is prominent in the vortex core region. 
At low field, the $d_{x^2-y^2}$-wave nature scarcely affects the
structure of the outer region of the core. 
With increasing field, the core region occupies a larger area of the unit
cell in the vortex lattice. 
Then, the $d_{x^2-y^2}$-wave nature gradually affects 
the whole structure of the vortex lattice. 
It affects the form factor of $h({\bf r})$ and the magnetic 
field distribution function. 
The former is detected by the SANS experiment, 
and the latter corresponds to the
resonance line shape in the $\mu$SR and NMR experiments. 
At higher field, we also have to consider the vortex lattice effect 
on the vortex structure. 
The orientation of the vortex lattice (i.e., the relative angle
between the $a$ axis and the NN vortex direction) is important for the
vortex structure. 

The vortex core radius decreases similarly for both the $s$-wave and 
the $d_{x^2-y^2}$-wave pairing  with increasing field. 
In the triangular vortex lattice case, 
the six NN directions are not equivalent in the vortex structure 
for the $d_{x^2-y^2}$-wave pairing. 
This may be the origin of the distortion from the triangular lattice. 
We showed the influence of the $d_{x^2-y^2}$-wave nature 
and the vortex lattice effect on the LDOS, 
and discussed the field dependence of the spatially averaged DOS. 
The spatial structure of the induced $s$-wave and $d_{xy}$-wave
order parameter in the $d_{x^2-y^2}$-wave superconductor was also presented. 
We note that the effect of the induced order
parameter was not included in our discussion of the $d_{x^2-y^2}$-wave
nature in Secs. \ref{sec:configulation}-\ref{sec:dos}. 
The $d_{x^2-y^2}$-wave nature appears even when the induced order
parameter of the other symmetry is absent.

The $d_{x^2-y^2}$-wave nature of the vortex structure $\Delta({\bf r})$, 
${\bf J}({\bf r})$ and $h({\bf r})$ was also studied in the GL theory
or the London theory by including nonlocal correction terms. 
We note that the GL theory only applies by the exact derivation 
near $T_c$.\cite{Werthamer} 
For other cases, we have to include nonlocal correction terms. 
We usually consider only the fourth order derivative terms as non-local 
terms, which contribute to the vortex structure in the correction of the
order $\ln(T_c/T)$. 
But, at lower temperature, we have to include higher order derivative terms. 
Their contribution is a higher order of $\ln(T_c/T)$. 
Further, as we increase the amplitude of the order parameter bellow 
$H_{c2}$, we also have to consider higher order nonlinear terms of
the order parameter than the usual $|\Delta|^2 \Delta$ term. 
It is difficult to consider all contributions of these higher order
terms by modifying the GL theory. 
The quasi-classical Eilenberger theory automatically includes all  
contributions of the higher order terms. 
Nevertheless, the $d_{x^2-y^2}$-wave nature and the vortex lattice effect 
obtained by the quasi-classical calculation  can be reproduced 
qualitatively by the GL theory with the correction of the fourth order 
derivative terms.\cite{IchiokaGL,Enomoto}  
We note that the $d_{x^2-y^2}$-wave nature in the resonance line shape $P(h)$
is also reproduced.\cite{IchiokaGL} 
This means that the extended GL theory is still useful as a
phenomenological theory for the qualitative study of the
$d_{x^2-y^2}$-wave nature in the vortex structure. 
The validity of the study with the nonlocal correction term 
(i.e., fourth order derivative term) in the GL theory or 
London theory\cite{Kogan,FranzL,Amin} can be 
confirmed by a comparison with the result obtained in a  
microscopic theory such as the quasi-classical theory.  
But, for a quantitative study, we have to use a microscopic theory.

In the London theory, the amplitude of the pair potential is treated 
as a constant outside the small region of the vortex core. 
However, this is only justifiable if the field is low enough. 
In our calculation where $T/T_c=0.5$, $|\Delta({\bf r})|$ shows a 
spatial variation at every region of a unit cell for $H>0.1H_{c2}$. 
We have to carefully use an approximation of the constant
amplitude over a  wide range of the applied field. 

We expect that the $d_{x^2-y^2}$-wave nature presented in this paper 
will be observed in experiments on the $d_{x^2-y^2}$-wave superconductors 
such as the high $T_c$ superconductors and 
the organic superconductors.\cite{Nakazawa} 
We recommend that the vortex structure be investigated into 
the higher field range, where the $d_{x^2-y^2}$-wave nature is eminent. 
At lower fields, we also consider the effect of the gradual transformation  
from the triangular to the square vortex lattice. 
Borocarbide superconductors such as ${\rm LuNi_2B_2C}$ are also good 
candidates for observing the $d_{x^2-y^2}$-wave nature.\cite{Nohara} 
These materials are considered to be $d_{x^2-y^2}$-wave
superconductors or $s$-wave superconductors with a highly anisotropic
Fermi surface with fourfold symmetry. 
Even for the latter $s$-wave case, we obtain a similar vortex
structure to the $d_{x^2-y^2}$-wave pairing case by the 
${\bf k}$-dependence of ${\bf v}_{\rm F}$ instead of the ${\bf k}$-dependence 
of $\phi(\theta)$.\cite{Hayashi}  
There, the ${\bf k}$-direction of large $v_{\rm F}$ plays the same role as
the small $|\phi(\theta)|$ direction. 
In experiments on ${\rm LuNi_2B_2C}$, the fourfold-symmetric behavior 
of $H_{c2}$ is reported when the field direction is rotated within the
$ab$ plane.\cite{Metlushko} 
From the $H_{c2}$ anisotropy, the magnitude of the
$d_{x^2-y^2}$-wave-{\it like} behavior is estimated as 
$\langle \phi(\theta)^2 (v_{{\rm F}x}^4 - 3 v_{{\rm F}x}^2 v_{{\rm F}y}^2)
\rangle_{\hat {\bf k}}/ 
4 \langle \phi(\theta)^2 (v_{{\rm F}x}^2+v_{{\rm F}y}^2 )
\rangle_{\hat {\bf k}}^2$=0.43, 
where $\langle\cdots\rangle_{\hat {\bf k}}$ means the
average on the Fermi surface.  
This magnitude is comparable to 0.25 of the $d_{x^2-y^2}$-wave
pairing case ($\phi(\theta)=\sqrt{2}\cos 2\theta$ and 
an isotropic 2D Fermi surface). 
So, the large $d_{x^2-y^2}$-wave-like anisotropy effect is expected in
the vortex structure. 
The gradual transition from the triangular to the square vortex lattice 
is also observed with increasing field.\cite{DeWilde,Eskildsen}  
As $\mu$SR and SANS experiments have begun on these materials, 
we expect that the $d_{x^2-y^2}$-wave-like nature will be detected there.

For the other symmetry of the anisotropic superconductors such as heavy
fermion superconductors or ${\rm Sr_2 Ru O_4}$, if the ${\bf k}$-dependence
of their pair potential or the Fermi surface structure has a large anisotropy
in the directions perpendicular to the applied magnetic field, the
anisotropy effect appears by the same origin as that discussed in this paper. 
It may give important information for determining the symmetry of
the order parameter.  

\section*{Acknowledgments} 

We would like to thank N. Hayashi and T. Sugiyama for their helpful
comments.  
Some of the numerical calculations in this work were carried out at the
Yukawa Institute Computer Facility, Kyoto University, and 
the Supercomputer Center of the Institute for Solid State Physics, 
University of Tokyo. 
One of the authors (M.I.) is supported financially by 
the Japan Society for the Promotion of Science.



\widetext

\newpage
\onecolumn

\leavevmode
\epsfxsize=6.0cm
\epsfbox{figp09.epsi}
\vspace{0.2cm}

\end{document}